\RequirePackage{lineno}
\documentclass[preprintnumbers,groupedaddress,superscriptaddress,amsmath,amssymb,longbibliography]{revtex4-1}
\usepackage{array}
\usepackage{bbm}
\usepackage{bm}
\usepackage{graphicx,graphics}
\usepackage{soul}
\usepackage[makeroom]{cancel}
\usepackage{braket}
\usepackage[linesnumbered,ruled,vlined]{algorithm2e}
\usepackage{mathtools}
\usepackage{multirow}
\usepackage{tikz-cd}
\DeclarePairedDelimiter\ceil{\lceil}{\rceil}
\usepackage[normalem]{ulem} 
\usepackage{xcolor} 
\usepackage{cleveref}
\usepackage{amsmath}
\usepackage{amsthm}

\newcommand{\rS}{\mathrm{S}}
\newcommand{\rE}{\mathrm{E}}
\newcommand{\rSE}{\mathrm{SE}}

\theoremstyle{definition}
\newtheorem{theorem}{Theorem}

\newtheorem{criterion}[theorem]{Criterion}

\begin{document}

\title{Probing non-Markovian quantum dynamics with data-driven analysis: \\ Beyond ``black-box'' machine-learning models}

\author{I. A. Luchnikov}
\email{luchnikovilya@gmail.com}
\affiliation{Russian Quantum Center, Skolkovo, Moscow 143025, Russia}
\affiliation{Skolkovo Institute of Science and Technology, Moscow 121205, Russia}
\affiliation{National University of Science and Technology ``MISIS”, Moscow 119049, Russia}

\author{E. O. Kiktenko}
\affiliation{Russian Quantum Center, Skolkovo, Moscow 143025, Russia}
\affiliation{Moscow Institute of Physics and Technology,
Moscow Region 141700, Russia}
\affiliation{Department of Mathematical Methods for Quantum Technologies,
Steklov Mathematical Institute of Russian Academy of Sciences, Moscow 119991, Russia}

\author{M. A. Gavreev}
\affiliation{Russian Quantum Center, Skolkovo, Moscow 143025, Russia}
\affiliation{Moscow Institute of Physics and Technology,
Moscow Region 141700, Russia}
\affiliation{National University of Science and Technology ``MISIS”, Moscow 119049, Russia}

\author{H. Ouerdane}
\affiliation{Skolkovo Institute of Science and Technology, Moscow 121205, Russia}

\author{S. N. Filippov}
\affiliation{Moscow Institute of Physics and Technology,
Moscow Region 141700, Russia} \affiliation{Department of Mathematical Methods for Quantum Technologies,
Steklov Mathematical Institute of Russian Academy of Sciences, Moscow 119991, Russia}
\affiliation{Valiev Institute of Physics and Technology of Russian Academy of Sciences, Moscow 117218, Russia}

\author{A. K. Fedorov}
\email{akf@rqc.ru}
\affiliation{Russian Quantum Center, Skolkovo, Moscow 143025, Russia}
\affiliation{Moscow Institute of Physics and Technology,
Moscow Region 141700, Russia}
\affiliation{National University of Science and Technology ``MISIS”, Moscow 119049, Russia}

\begin{abstract}
A precise understanding of the influence of a quantum system's environment on its dynamics, which is at the heart of the theory of open quantum systems, is crucial for further progress in the development of controllable large-scale quantum systems. However, existing approaches to account for complex system-environment interaction in the presence of memory effects are either based on heuristic and oversimplified principles or give rise to computational difficulties. In practice, one can leverage on available experimental data and replace first-principles simulations with a data-driven analysis that is often much simpler. Inspired by recent advances in data analysis and machine learning, we propose a data-driven approach to the analysis of the non-Markovian dynamics of open quantum systems. Our method allows, on the one hand, capturing the most important characteristics of open quantum systems such as the effective dimension of the environment and the spectrum of the joint system-environment quantum dynamics, and, on the other hand, reconstructing a predictive model of non-Markovian quantum dynamics, and denoising the measured quantum trajectories. We demonstrate the performance of the proposed approach with various models of open quantum systems, including a qubit coupled with a finite environment, a spin-boson model, and the damped Jaynes-Cummings model.
\end{abstract}

\maketitle

\section{Introduction}

Understanding and predicting dynamics of quantum many-body systems is a formidable challenge, and an essential step towards the development of quantum computing devices~\cite{ladd2010quantum}. Recent experiments~\cite{georgescu2014quantum, bloch2012quantum, blatt2012quantum, friis2018observation, trotzky2012probing, mazurenko2017cold, keesling2019quantum, barreiro2011open} have demonstrated the realization of quantum computing and simulation protocols with quantum many-body systems of an intermediate scale (dozens of particles). Nevertheless, even with the achieved outstanding level of control in these experiments it remains challenging to isolate a quantum system from the environment~\cite{barreiro2011open}, which is the main source of decoherence causing errors. This fact significantly limits applications of such systems for quantum information processing and prevents the full experimental verification of various effects related to the nonequilibrium dynamics \cite{vasseur2016nonequilibrium, RevModPhys.91.021001, bernien2017probing}.

The theory of open quantum systems ~\cite{breuer2002theory, de2017dynamics, alicki2007quantum, accardi2013quantum} aims to provide a complete description of such nonequilibrium dynamical effects at the quantum level. Within such a framework, one can arbitrarily partition a quantum many-body system and its environment and describe each part as an individual open system. This powerful idea allows studying a nonequilibrium many-body system by considering smaller parts of it; for example, an individual qubit in a large ensemble of qubits interacting with each other and their environment. At first sight, dealing with one qubit instead of an ensemble of entangled qubits seems to be a much simpler problem, because it does not suffer from the computational difficulties raised by the exponentially large Hilbert space in the number of subsystems. Indeed, the dynamics of the subsystem's density matrix (e.g. one qubit density matrix) can be described exactly by the Nakajima-Zwanzig equation~\cite{nakajima1958quantum, zwanzig1960ensemble}, which contains a convolution (memory) kernel. Time-convolution is necessary to take into account the non-Markovianity of the general open quantum dynamics. Non-Markovianity, i.e. effects of memory, are among the most challenging problems to solve for the proper description of the general open quantum system~\cite{de2017dynamics}. The problem is that the exact derivation of the Nakajima-Zwanzig equation is of the same complexity as the calculation of the coupled system and environment quantum dynamics~\cite{breuer2002theory}. 

This problem can be approached by considering different approximations such as the Born-Markov approximation~\cite{moy1999born, accardi2013quantum} or simplified models of open quantum dynamics that are exactly solvable~\cite{unruh1995maintaining, palma1996quantum, vacchini2010exact, teretenkov2019non}. However, the class of problems that can be studied analytically is limited to a handful of exceptional situations. An alternative approach is to use numerical techniques such as the non-Markovian quantum state diffusion~\cite{piilo2008non, diosi1997non, strunz1999open, diosi1998non}, the hierarchical equations of motion~\cite{tanimura1990nonperturbative, tanimura1989time}, the time-evolving matrix product operators~\cite{strathearn2018efficient}, the method based on optimized auxiliary oscillators~\cite{mascherpa2020optimized}, the dressed quantum trajectories method~\cite{polyakov2019dressed}, the Dirac-Frenkel time-dependent variational approach with the Davydov ansatz~\cite{wang2017finite}, or the time-evolving density with orthogonal polynomials algorithm~\cite{chin2010exact, prior2010efficient}, just to name a few. At the same time, open quantum systems studied under real experimental conditions are usually too complex and cannot always be described with existing numerical methods developed for illustrative models. For instance, most of the numerical methods, including those listed above, are devoted to simulating systems with an environment made of non-interacting quantum oscillators~\cite{de2017dynamics} or non-interacting fermions~\cite{schon1990quantum, anderson1961localized}, so they are unable to simulate systems with more complex environments, such as the spin environment~\cite{saykin2002relaxation, hanson2008coherent, fischer2007correlated, lerose2021influence, sonner2021influence}. Moreover, since in a real experimental setting the joint Hamiltonian of a system and its environment is often unknown, exhaustive spectroscopy experiments are required to recover the Hamiltonian parameters and simulate the actual system's dynamics at the microscopic level~\cite{acosta2009high, soshenko2018microwave}.

Since building a precise analytical or numerical description of open quantum dynamics from a microscopic model is a notoriously difficult problem in the general case, one can try to build a precise data-driven model of open quantum dynamics using experimentally measured data. In addition, data-driven approaches do not require reconstruction of the joint system and environment microscopic model. A number of data-driven techniques for the analysis of the dynamics of open systems have been recently proposed. For example, Ref.~\cite{cerrillo2014non} introduces a method for a data-driven reconstruction of the discrete-time Nakajima–Zwanzig equation, namely the Transfer-Tensor method (TTM) \cite{buser2017initial, rosenbach2016efficient, kananenka2016accurate, gelzinis2017applicability, pollock2018tomographically, chen2020non, jorgensen2019exploiting, jorgensen2020discrete, gherardini2021transfer}, that can be used for dynamics prediction. In Ref.~\cite{banchi2018modelling} a recurrent neural network based method~\cite{salehinejad2017recent} for data-driven identification of open quantum dynamics has been developed. Although these methods provide efficient predictive models, they serve as ``black boxes'', i.e. they do not allow one to unravel the physical picture of the underlying processes~\cite{iten2020discovering, raissi2019physics, lusch2018deep}.

\begin{figure}[ht]
    \centering
    \includegraphics[width=\linewidth]{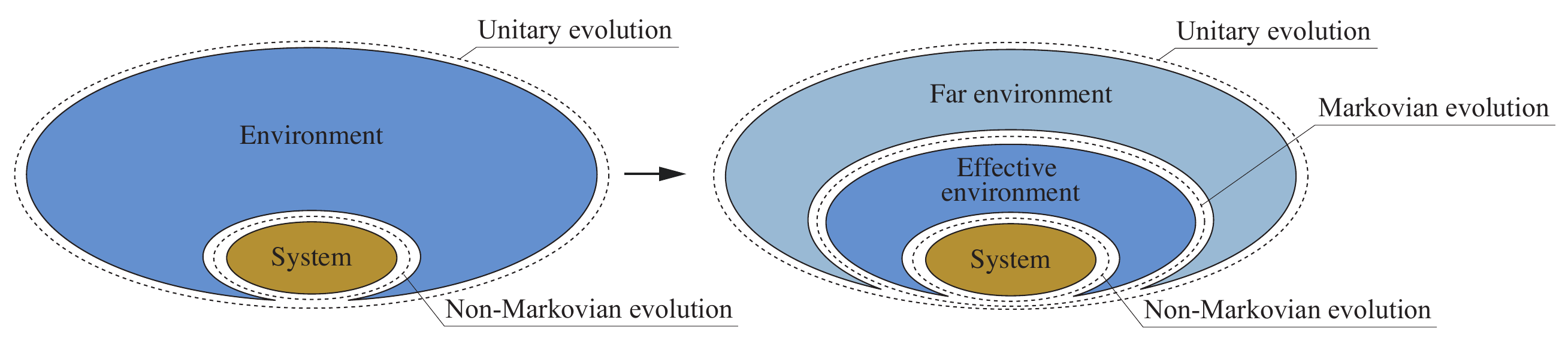}
    \caption{
    Physical intuition underpinning the Markovian embedding for the non-Markovian dynamics of the finite-dimensional system interacting with a high-dimensional (possibly infinite-dimensional) environment.
    One can split the system's environment into two parts: The finite-dimensional effective environment, which is responsible for memory effects in the system's dynamics, and the remaining far environment responsible for dissipation. 
    It allows one to consider a Markovian dissipative dynamics of the finite-dimensional extended system consisting of the original system and its effective environment.}
    \label{fig:markovian_embedding_interpretation}
\end{figure}

One of the most natural ways to describe open quantum dynamics is embedding the non-Markovian system dynamics into a Markovian dynamics of the system and the effective environment of a finite dimension~\cite{budini2013embedding}. The physical intuition behind such embedding is summarized in Fig.~\ref{fig:markovian_embedding_interpretation}. One can think of the environment that induces the system's non-Markovian dynamics as a two-component system consisting of effective and far environments. The effective environment is responsible for memory effects: some information about the system is recorded into the effective environment and then affects the system at later time.
The dimension of the effective environment determines the complexity of the non-Markovian system's dynamics. In contrast to the effective environment, the interaction with the far environment does not entail a backflow of information, and leads to a purely dissipative dynamics. In this way, the system and the effective environment instantiate a Markovian embedding, and their collective dynamics is both dissipative and Markovian. We emphasize that Markovian embedding is not just a ``black box'' model of open quantum dynamics since it provides insights into the properties of both the system and its environment. Moreover, Markovian embedding can be reconstructed with only information about the non-Markovian dynamics of a system, which is accessible in real experimental conditions~\cite{luchnikov2020machine, guo2020tensor}. 

In this work, we harness data-driven Markovian embedding reconstruction, and present a complete framework for general open quantum dynamics analysis. Our framework can be seen as a combination of the alternative ``data-friendly'' view on non-Markovian quantum dynamics, that we develop in first sections, and linear machine-learning methods~\cite{gavish2014optimal, schmid2010dynamic, tu2013dynamic, bishop2006pattern}. Input data include trajectories of a system at discrete time steps, a guess of the memory depth of the non-Markovian process, and the level of noise generated during experimental reconstruction of trajectories. As output, our framework returns the predictive model of non-Markovian quantum dynamics, the dimension of the effective environment~\cite{luchnikov2019simulation, shrapnel2018quantum, guo2020tensor}, the eigenfrequencies of the joint system and environment quantum dynamics, and the denoised trajectories of the system as a valuable byproduct. Note that our framework has the distinct advantage to rely on the linear methods of machine learning, which are known to be scalable, data efficient and yield the exact solution. We illustrate the performance of our scheme with several paradigmatic examples of realistic models such as a qubit coupled with a finite-dimensional environment, a spin-boson model, and the damped Jaynes-Cummings model.

\section{Time-delay Embedding of non-Markovian quantum dynamics}
\label{seq:theory}
\subsection{Quantum trajectories' Markovian dynamics}
In this section we discuss an alternative view on non-Markovian open quantum dynamics that is the basis of the proposed data processing scheme. The alternative lies in turning from the density matrix based description of non-Markovian quantum dynamics to quantum trajectories based description, where a quantum trajectory is seen as a concatenation of several subsequent in time density matrices. This description is less physics-motivated and more suitable for the data-driven analysis we propose. Consider a $d$-dimensional quantum system undergoing non-Markovian dynamics \cite{milz2020quantum} due to the interaction with a $d_{\rE}$-dimensional environment. Throughout the paper, we assume that the Hamiltonian driving dynamics of the system and environment as a whole remains time-independent. In the development that follows, $d$ and $d_{\rE}$ are finite, though the environment may consist of a large number of subsystems (particles), so one can have $d_{\rE}$ exponentially large in the number of particles. Assume one has access to the system states (density matrices) $\varrho(t)\in\mathbb{C}^{d\times d}$ at consecutive and equidistant time instants $t=0,1,2,\ldots$, separated by the constant experimental time resolution $\tau$, and one is able to reconstruct $\varrho(t)$ using quantum state tomography. Throughout the paper we use $\tau$ as time units, i.e. the physical time is expressed as $\tau t$.
Performing the tomographic experiment multiple times, one reconstructs $K$-element sequences of the system's states:
\begin{equation}
    T_K(t) := (\varrho(t), \varrho(t+1), \ldots, \varrho(t+K))\in\mathbb{C}^{K\times d\times d},
\end{equation}
which we refer to as system \emph{trajectories}. Assuming that the observed non-Markovian dynamics has a finite memory depth, for a sufficiently large value of $K$, there exists a linear map $M$ that connects two subsequent trajectories through a Markovian master equation:
\begin{equation} \label{eq:MarkovianM}
    T_K(t+1) = M[T_K(t)].
\end{equation}
The rigorous criteria for $K$ to be sufficient (existence of $M$) is given in Appendix \ref{appendix_1}. One can think of $K$ as the number of previous in-time density matrices of the system, which is enough to determine the next in-time state of the system. Note also that while one can express the physical memory depth as $\tau K$, which is a more fundamental quantity, $K$ itself is more suitable for data-driven analysis. In the Appendix \ref{appendix_1} we also show that for a finite $d_\rE$ one always has a finite sufficient $K$ that does not exceed $d^2d_\rE^2$. Note, however, that in the worst case, the minimal sufficient $K$ scales with the number of environment's subsystems (particles) exponentially and Eq.~\eqref{eq:MarkovianM} quickly becomes intractable for a numerical analysis. Our conjecture is that this worst case corresponds to a dynamical chaos regime, as we discussed in more detail in Appendix \ref{appendix_1}, for which long-time dynamics prediction is impossible. Although, one can use the data-driven analysis (discussed in the next section) to identify chaos in the system's dynamics, our scheme is suitable only for finite-memory (non-chaotic) non-Markovian dynamics. In what follows, $K$ is always assumed to be sufficient. We also refer to a sufficient $K$ as to a {\it memory length}.

\begin{figure}[h]
    \centering
    \includegraphics[width=0.7\linewidth]{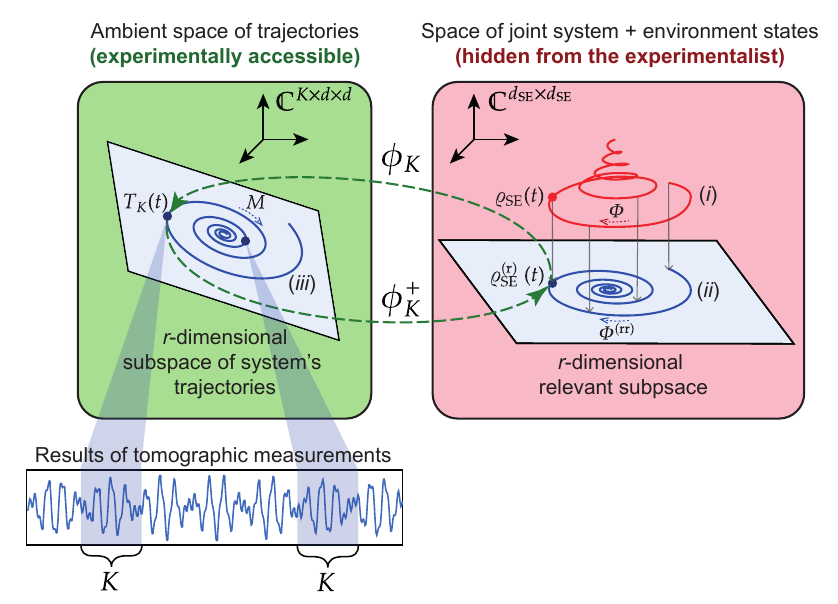}
    \caption{Connection between the joint dynamics of the system and its environment with the observed dynamics of trajectories. The curve (\emph{i}) sketches the joint dynamics of the system and its environment $\varrho_\rSE(t)$ driven by the quantum channel $\Phi$. The curve (\emph{ii}) sketches the joint dynamics of the system and the effective environment $\varrho_\rSE^{(r)}(t)$ driven by the map $\Phi^{\rm (rr)}$. The curve (\emph{iii}) sketches the observed trajectories dynamics $T_K(t)$ driven by the map $M$. There is the one-to-one correspondence between curves (\emph{ii}) and (\emph{iii}) that is described by the linear map $\phi_K$. Both dynamics (\emph{ii}) and (\emph{iii}) are Markovian, their Master equations are connected via $\phi_K$ and they can be seen as `shadows' of the dynamics (\emph{i}).}
    \label{fig:embedding_illustration}
\end{figure}

The general idea of considering the Markovian dynamics of $T_K(t)$ instead of the non-Markovian dynamics of $\varrho(t)$ is also known as a \emph{time-delay embedding} and it is routinely employed across wide variety of contexts, including, Koopman operators \cite{arbabi2017ergodic, arbabi2017study, kamb2020time, brunton2017chaos}, closure modeling \cite{pan2018data}, time series modeling \cite{chen1989representations, hegger1999practical} and reinforcement learning \cite{hausknecht2015deep}.
One can also think of the time-delay embedding as a particular way to construct a \emph{Markovian embedding} \cite{bennink2019quantum, campbell2018system, xue2019modeling, xue2015quantum, budini2013embedding, luchnikov2019simulation, luchnikov2020machine} of non-Markovian quantum dynamics. The framework being developed here also seems very close in spirit to the TTM. However, there is a substantial difference. The TTM operates with a map from the space of trajectories to the space of density matrices, while here we operate with a map from the space of trajectories to the same space. This framework turns out to be more fruitful since it allows us to study the environment properties in addition to the system's dynamics prediction. In particular we show that the observed trajectories dynamics is the same, up to linear rescaling, as joint dynamics of the system and so called effective part of the environment. This finding is illustrated in Fig.~\ref{fig:embedding_illustration}.
The given figure and the concepts involved are discussed in detail below. Since the main distinguishing feature of the presented method is its ability to explore the physics of the environment, the central question we address in this section is: what knowledge about the system and its environment causing non-Markovianity one can extract from the analysis of the Markovian dynamics of system's trajectories?

\subsection{Unraveling the joint system and environment dynamics via the trajectories' dynamics}
To start addressing this question, let us first note that states  $\varrho(t)$ can be expressed via the corresponding joint system and environment states $\varrho_\rSE(t)\in\mathbb{C}^{d_{\rSE}\times d_{\rSE}}$ as follows
\begin{equation}\label{eq:partial-trace}
    \varrho(t)={\rm Tr}_\rE(\varrho_\rSE(t)),
\end{equation} 
where $d_{\rSE}:=dd_{\rE}$ and ${\rm Tr}_\rE$ denotes a partial trace over the environment. Let us also introduce a quantum channel $\Phi$ describing the discrete-in-time evolution of the joint states:
\begin{equation} \label{eq:jointPhi}
    \varrho_\rSE(t+1) = \Phi[\varrho_\rSE(t)].
\end{equation}
Since ${\rm Tr}_\rE$ is a linear map and the dynamics of $\varrho_\rSE(t)$ is linear as well, one can introduce a linear map $\phi_K$ returning $T_K(t)$ from $\varrho_{\rSE}(t)$:
\begin{equation}\label{eq:phi_K}
    T_K(t) = \phi_K [\varrho_{\rSE}(t)].
\end{equation}
The precise definition of $\phi_K$ is given in the Appendix~\ref{appendix_1}; for the purposes of this section we only need the linearity property of this map.

The analysis of $\phi_K$ provides insights into the connection between the system trajectories' dynamics and the actual joint system and environment dynamics. It turns out that the linear map $\phi_K$ defines which `part' of the joint system and environment dynamics affects the observed trajectories. Let us call the {\it irrelevant (relevant) subspace} the kernel (the support) of $\phi_K$. The corresponding projectors take the form:
\begin{equation}
    \pi^{\rm (i)}_K = {\rm Id} - \phi_K^+\phi_K, \quad \pi^{\rm (r)}_K = \phi_K^+\phi_K,
\end{equation}
where ${\rm Id}$ denotes the identity map, $\phi_K^+$ is the Moore–Penrose inverse of $\phi_K$, and superscripts ${\rm (i)}$ and ${\rm (r)}$ stand for irrelevant and relevant respectively. These projectors allow us to introduce the `irrelevant' and `relevant' parts of the joint state that read
\begin{equation}
    \varrho_\rSE^{\rm (i)}(t) = \pi^{\rm (i)}_K[\varrho_\rSE(t)],\quad
    \varrho_\rSE^{\rm (r)}(t) = \pi^{\rm (r)}_K[\varrho_\rSE(t)].
\end{equation}
Since $\varrho_\rSE^{\rm (i)}(t)$ belongs to the kernel of $\phi_K$, only $\varrho_\rSE^{\rm (r)}(t)$ defines $T_K(t)$. Therefore $\varrho_\rSE^{\rm (i)}(t)$ and $\varrho_\rSE^{\rm (r)}(t) $ have a clear operational meaning: The dynamics of $\varrho_\rSE^{\rm (i)}(t)$ is hidden, while $\varrho_\rSE^{\rm (r)}(t)$ can be considered as a joint state of the system and the `effective' environment contributing to the observed non-Markovian dynamics of the system. Note, however, that $\varrho_\rSE^{\rm (i)}(t)$ and $\varrho_\rSE^{\rm (r)}(t)$ no longer have a unit trace. Moreover, there is no guarantee that they are positive semidefinite or even Hermitian. One can think of the projection on the relevant subspace as an analog of the partial trace over those degrees of freedom of the environment that do not contribute to the memory effects. We note that the similar projections on the relevant and irrelevant subspaces, yet with different meanings, are employed in the Nakajima-Zwanzig projection method \cite{breuer2002theory}.

An important characteristic of the linear map $\phi_K$ is the rank $r$ of its matrix.
It provides the dimension of the relevant subspace. In the next subsection, we show that the dynamics in the relevant subspace is independent of the dynamics in the irrelevant subspace. So, to fully describe the non-Markovian dynamics of the system, it is enough to consider only the relevant subspace that can be seen now as the space of the joint system and effective environment density matrices.
By the construction, $d_\rE^{\rm eff}$ provides the minimal dimensionality of the environment that can describe the non-Markovian behavior of the system.
This observation allows us to introduce the dimension of the effective environment through the following relation $r = d^2[d^{\rm eff}_\rE]^2$, which simply states that the joint system and effective environment density matrix has size $dd^{\rm eff}_\rE \times dd^{\rm eff}_\rE$. Note, however, that there is no guarantee that $d^{\rm eff}_\rE = \sqrt{r / d^2}$ results in an integer number, since $r$ can be an arbitrary integer. So, to make $d^{\rm eff}_\rE$ an integer, one can either round it up or down. Rounding down contradicts the fact that $\sqrt{r / d^2}$ is an underestimation of $d^{\rm eff}_\rE$ since some of the degrees of freedom of the environment may not affect the system dynamics. Therefore, one needs to round $\sqrt{r / d^2}$ up, or in other words one needs to complement the relevant subspace with additional degrees of freedom that, however, do not affect the dynamics of the system. Clearly, this can always be done and not in a unique way. We use this trick only here and for the sole purpose to justify the following final estimation of the effective environment dimension that reads:
\begin{equation}
    \label{eq:d_eff_estimation}
    d_\rE^{\rm eff} := \ceil*{\sqrt{r / d^2}},
\end{equation}
where $\ceil*{\cdot}$ stands for the rounding up operation. Note that the value of $r$ and $d^{\rm eff}_\rE$ are independent of $K$ because $K$ is sufficient (see Appendix \ref{appendix_1}). The concept of the effective environment's dimension has been actively studied in recent literature~\cite{shrapnel2018quantum, pollock2018non, pollock2018operational, luchnikov2019simulation, luchnikov2020machine} and has been shown to be an important characteristic of the environment. One can think of $d_\rE^{\rm eff}$ as a quantitative indicator of non-Markovianity, the dimensionality of the memory, or the complexity of non-Markovian dynamics. We remark that an exponentially large $d_\rE^{\rm eff}$ corresponds to utterly complex dynamics / dynamical chaos.

\subsection{Equivalence between the trajectories' dynamics and the dynamics of the system and its effective environment}
Next, consider the $r$-dimensional subspace ${\cal C}_K\equiv {\rm Im}(\phi_K)  \subseteq {\mathbb{C}^{K\times d\times d}}$, where ${\rm Im}$ stands for the image of $\phi_K$. All experimentally accessible trajectories $T_K(t)$ lie in this subspace by definition; therefore, we refer to it as {\it trajectories subspace}. Note that $\phi_K$ is the linear bijection between the relevant subspace and the trajectories subspace. The Moore–Penrose inverse $\phi^+$ inverts this bijection. The bijection between this subspaces is complemented by the following identity between $\Phi$, $M$, and $\phi_K$:
\begin{equation} \label{eq:dynamics_isomorphism}
    M \phi_K = \phi_K \Phi.
\end{equation}

One can think about Eq.~\eqref{eq:dynamics_isomorphism} as about the following commutative diagram: 

\begin{center}
\begin{tikzcd}
    \varrho_\rSE(t) \arrow[r, "\Phi"] \arrow[d, "\phi_K"]
    & \varrho_\rSE(t+1) \arrow[d, "\phi_K"] \\
    T_K(t) \arrow[r, "M"]
    & T_K(t+1)
\end{tikzcd}.     
\end{center}

It says that one can first make a time step forward in the space of the joint system and environment states and then go to the corresponding trajectory via $\phi_K$ or first go to a trajectory via $\phi_K$ from a given join state and then make a time step forward by applying $M$ in the trajectories subspace.
Note that the identity Eq.~\eqref{eq:dynamics_isomorphism} follows from the fact that $M$ describes the temporal evolution of any trajectory. The existence of the bijection between the relevant subspace and ${\cal C}_K$ and Eq.~\eqref{eq:dynamics_isomorphism} mean that the trajectories' dynamics and the joint system and effective environment dynamics are identical in a sense that is explained below. Let us decompose $\Phi$ into four terms
\begin{equation}\label{eq:Phi-decomp}
    \Phi = \Phi^{\rm(ii)} + \Phi^{\rm(ir)} + \Phi^{\rm(ri)} + \Phi^{\rm(rr)},
\end{equation}
where
$\Phi^{\rm(xy)} := \pi_K^{\rm (x)}  \Phi  \pi_K^{\rm (y)}$, ${\rm x}, {\rm y} \in \{{\rm i}, {\rm r}\}$.
One can see that each term has a clear operational meaning: $\Phi^{\rm (rr)}$ and $\Phi^{\rm (ii)}$ govern the dynamics in the relevant and irrelevant subspace respectively, while $\Phi^{\rm (ir)}$ and $\Phi^{\rm (ri)}$ correspond to `cross-flows' between the subspaces. Note, however, that $\Phi^{\rm(xy)}$ are neither trace-preserving nor completely positive maps in general. Let us substitute the decomposition Eq.~\eqref{eq:Phi-decomp} in Eq.~\eqref{eq:dynamics_isomorphism}
\begin{equation}
    M \phi_K = \phi_K  \Phi^{\rm(ii)} +  \phi_K \Phi^{\rm(ir)} + \phi_K \Phi^{\rm(ri)} + \phi_K \Phi^{\rm(rr)}.
\end{equation}
One can compare supports of all terms of the identity above and identifies two linearly independent identities that read 
\begin{eqnarray}
    &&M\phi_K = \phi_K \Phi^{({\rm ir})} + \phi_K \Phi^{({\rm rr})},\nonumber\\
    &&0 = \phi_K \Phi^{\rm (ii)} + \phi_K \Phi^{\rm (ri)}.
\end{eqnarray}
By noticing that $\phi_K \Phi^{\rm (ii)} = 0$ and $\phi_K \Phi^{\rm (ir)} = 0$ due to the fact that $\phi_K$ maps any vector from the irrelevant subspace to zero one finally gets
\begin{eqnarray}
    &&M\phi_K = \phi_K \Phi^{\rm (rr)},\nonumber\\
    &&0 = \phi_K \Phi^{\rm (ri)}.
\end{eqnarray}
The identity $0 = \phi_K \Phi^{\rm (ri)}$ implies the absence of `flow' from the irrelevant subspace to the relevant one. In other words $\Phi^{\rm (rr)}$ fully describes the dynamics in the relevant subspace:
\begin{equation} \label{eq:Markovian-evol-of-projs}
    \varrho_\rSE^{\rm (r)}(t+1) = \Phi^{\rm (rr)}\left[\varrho_\rSE^{\rm (r)}(t)\right],
\end{equation}
since there is no effect of the irrelevant subspace on the dynamics in the relevant subspace. Moreover, one can see that Eq.~\eqref{eq:Markovian-evol-of-projs} is a Markovian master equation. Let us multiply the identity $M\phi_K = \phi_K \Phi^{\rm (rr)}$ by $\phi_K^+$ on the right side
\begin{equation}
    M\phi_K\phi_K^+ = \phi_K \Phi^{\rm (rr)} \phi_K^+.
\end{equation}
Bearing in mind that the action of $M$ on the orthogonal complement of ${\cal C}_K$ does not have a physical meaning and can be set arbitrarily, we set it equal to zero. Hence, $M = M\phi_K\phi_K^+$, and we get 
\begin{equation}
    M = \phi_K \Phi^{\rm (rr)} \phi_K^+.
\end{equation}
Now we can claim that the dynamics of trajectories and the joint dynamics of the system and its effective environment are identical up to a linear rescaling. Indeed, both dynamics are Markovian, there is the linear bijection between states, i.e. $T(t) = \phi_K\left[\varrho_\rSE^{\rm (r)}(t)\right]$ and the corresponding linear identity between master equations $M = \phi_K \Phi^{\rm (rr)} \phi_K^+$. 
This also implies that $M$ and $\Phi^{\rm (rr)}$ have completely the same eigenvalues and their left and right eigenvectors are connected via $\phi_K$. Moreover, since $\Phi$ has block triangular form w.r.t relevant and irrelevant subspaces ($\Phi^{\rm (ri)} = 0$), the set of eigenvalues of $\Phi$ includes all the eigenvalues of $\Phi^{\rm (rr)}$, and so the eigenvalues of $M$. The overall concept described above is illustrated in Fig.~\ref{fig:embedding_illustration}.

The fact that the dynamics of the system and environment factorizes into two parts: the ``relevant'' part, which is Markovian, and the independent ``irrelevant'' part, may seem artificial at first glance. Indeed, this property should imply some symmetry that is not inherent to an arbitrary system. In fact, the exact factorization in general, except for some trivial cases of two non-interacting subsystems or highly symmetric cases, results trivially in the ``relevant'' part that covers the entire state space and a one-dimensional irrelevant part. However, an approximate factorization (with some acceptable error) may be far more productive and lead to a significant dimensionality reduction of the ``relevant'' part compared to the entire state space. The data-driven approach we introduce later builds an approximate factorization by setting a threshold defined by the noise amplitude.

The main results of this section may be summarized as follows. If a non-Markovian dynamics has a finite memory length $K$ then: (I) The dynamics of trajectories is driven by the Markovian master equation~\eqref{eq:MarkovianM}; (II) All the experimentally accessible trajectories lie in a subspace of dimension $r={\rm rank}(M)$ (trajectories subspace) and form an $r$-dimensional Markovian embedding;
(III) There exists an $r$-dimensional relevant subspace of the joint system and environment density matrices' space. The dynamics in the relevant subspace is Markovian and identical to the dynamics of trajectories up to the bijective linear transformation $\phi_K$. The dynamics in the orthogonal complement of the relevant subspace does not affect dynamics of trajectories at all; (IV) One can think of the dynamics in the relevant subspace as the dynamics of the joint system and effective environment. One can estimate the dimension of the effective environment $d_\rE^{{\rm eff}}$ using Eq.~\eqref{eq:d_eff_estimation}. Both $r$ and $d_\rE^{{\rm eff}}$ are quantitative indicators of the non-Markovian dynamics complexity; (V) As a consequence of (III), all the eigenvalues of the dynamical maps $M$ and $\Phi^{\rm (rr)}$ are the same, and are part of the set of eigenvalues of $\Phi$.

These claims allow one to gain valuable knowledge about the system and its environment from the master equation describing the system's trajectories and the dimension of the subspace where these trajectories are embedded. The only missing ingredient at this stage, is a scheme to extract these objects from noisy measured trajectories.

\section{Algorithms for processing of quantum trajectories}
\label{seq:algorithms}

\subsection{Effective environment dimension identification and data denoising}
\begin{figure}[h]
    \centering
    \includegraphics[scale=0.65]{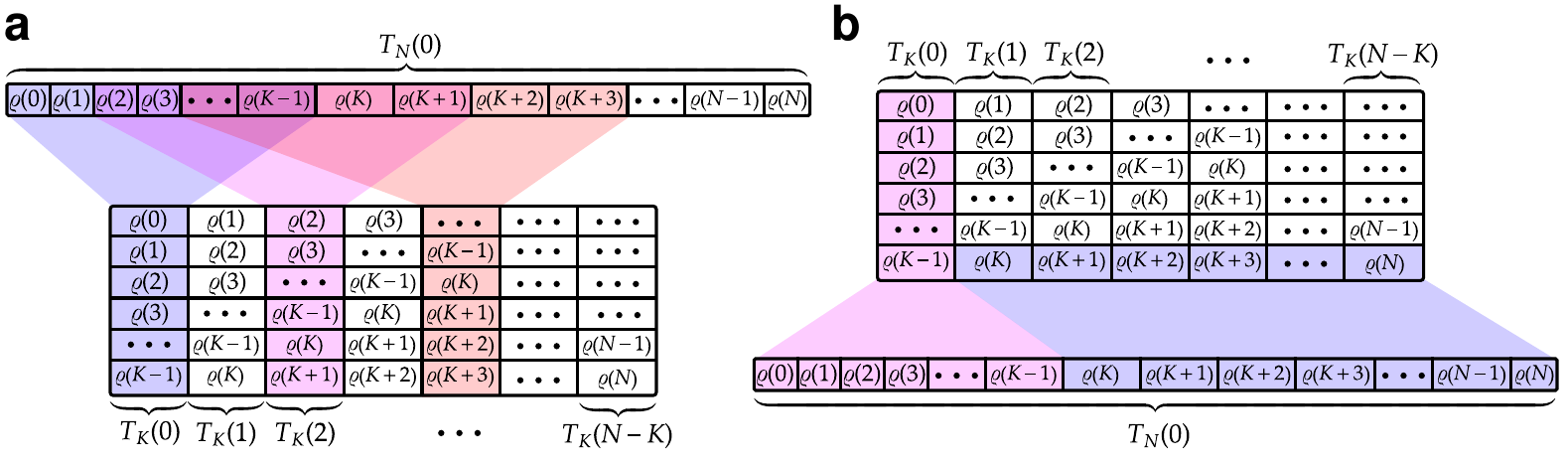}
    \caption{{\bf a} Graphical interpretation of slicing a trajectory into smaller chunks. {\bf b} Reconstruction of the whole trajectory of length $N$ back from chunks.}
    \label{fig:trajectory_slicing}
\end{figure}

As we discussed in the previous section, the only data assumed to be experimentally accessible are the system's discrete-in-time quantum trajectories. Those quantum trajectories can be reconstructed via quantum state tomography \cite{chuang1997prescription, altepeter2003ancilla}, i.e., for each trajectory, one runs multiple experiments with completely identical initial conditions and collects measurement outcomes statistics in order to reconstruct underlying density matrices in different moments of time. Since in real experiments each measurement takes some finite time $\tau_m$, one has a natural restriction $\tau \geq \tau_m$. Note also that the time resolution $\tau$ should be chosen small enough to capture all necessary short-time dynamical effects but not too small not to make an experiment lasts too long collecting too many data-points. Note, that subsequent experiment runs must be sufficiently separated in time, in order to eliminate memory effects from the previous run. Let us assume that for $L$ different initial states of the system the measured data set consists of quantum trajectories of length $N > K$, where $N$ is the total number of discrete time steps per each trajectory and $K$ is assumed to be sufficient for the Markovian dynamics of trajectories. This set reads
\begin{equation}
    {\cal S} = \left\{T_N^{(i)}(0)\right\}_{i=1}^{L},
\end{equation}
where $i$ enumerates initial states.
To prepare data for the analysis we construct their time-delay embedding. We slice each trajectory $T_N^{(i)}(0)$ into $N - K + 1$ smaller $K$-steps chunks $\{T^{(i)}_K(t)\}_{t=0}^{N-K}$ as it is shown in Fig. \ref{fig:trajectory_slicing}{\bf a}.
The resulting set of chunks $\left\{\{T_K^{(i)}(t)\}_{t=0}^{N-K}\right\}_{i=1}^L$ are processed further.

First, one can estimate the subspace ${\cal C}_K$ as the linear span of all the trajectories chunks, i.e.
\begin{equation}
    \label{eq:noiseles_ck_estimate}
    {\cal C}_K \approx {\rm span}\left(\left\{\{T_K^{(i)}(t)\}_{t=0}^{N-K}\right\}_{i=1}^L\right).
\end{equation}
The question arises: how many trajectories $L$ have to be measured in order to provide a reliable estimate of ${\cal C}_K$? For noiseless trajectories it turns out that in most cases one needs only one long enough trajectory. This can be explained as follows. The estimate Eq.~\eqref{eq:noiseles_ck_estimate} for only one trajectory reads
\begin{equation}
    {\cal C}_K \approx {\rm span}\left(\left\{T_K^{(1)}(t)\right\}_{t=0}^{N-K}\right) = {\rm span}\left(\left\{M^tT_K^{(1)}(0)\right\}_{t=0}^{N-K}\right).
\end{equation}
Note that unless the initial trajectory chunk $T_K^{(1)}(0)$ not in a low-dimensional invariant subspace of $M$, for $N-K = r$ one has ${\rm span}\left(\left\{M^tT_K^{(1)}(0)\right\}_{t=0}^{N-K}\right)$ covering the entire trajectories subspace. Therefore, only one trajectory is enough. This is also supported by the numerical experiments that are demonstrated further. For example, in Fig.~\ref{fig:dmd_vs_tt} one can see that the average error of prediction for quite a few numerical experiments with $L=1$ and different other parameters, including noise amplitude, is about $0.06$, which is not that much. However for highly symmetric systems it might be the case that ${\rm span}\left(\left\{M^tT_K^{(1)}(0)\right\}_{t=0}^{N-K}\right)$ covers only part of the trajectories subspace. In this case, one needs to sample trajectories with different initial states. It can be done in different ways. For example, before measuring a trajectory one can perturb the system in various ways, e.g. measure it on an arbitrary basis or apply some control signal to it. In our numerical experiments, we prepare a unique initial state as follows. We set the initial joint system and environment state in the stationary thermal state and then replace the system with a new one in some state of our choice. This helps us to explore the trajectories subspace better. To determine the value of $L$ in practice, one can sample a new trajectory with a unique initial state until the estimate Eq.~\eqref{eq:noiseles_ck_estimate} saturates.

All the trajectories bear some noise induced by finiteness of the measurements statistics, imperfections of measurement devices, etc. This results in an additive noise
\begin{equation}
    T^{(i)}_K(t) \longrightarrow T^{(i)}_K(t) + \delta T^{(i)}_K(t),
\end{equation}
where $\delta T^{(i)}_K(t)$ is a noise term. Note that this is not a noise induced by the interaction of the system with the environment that has an entirely different nature. The linear span of noisy trajectories always tends to the ambient space $\mathbb{C}^{d\times d \times K}$ with the growing number of trajectories in a data set and does not correspond to the genuine subspace ${\cal C}_K \subseteq \mathbb{C}^{K \times d\times d}$ of trajectories. Thus, one requires a method that identifies the genuine trajectories subspace in the presence of background noise. In other words, one needs to find a low-dimensional hyperplane of an appropriate dimension such that the data points (trajectories) lie as close to this hyperplane as possible. The dimension of the hyperplane is chosen to separate signal from noise. This is a typical machine learning task usually performed by using a principal component analysis (PCA) with automatic rank determination. We found the automatic rank determination procedure described in \cite{gavish2014optimal} the most suitable for our case. 
The overall algorithm is summarized here:

{\it Step 1.} One concatenates all the trajectories chunks into a matrix
\begin{equation}
    H = \underbrace{\begin{bmatrix} | & | & | & | & | & | & | & | & | & |\\ T_K^{(1)}(0) & \dots & T_K^{(L)}(0) & T_K^{(1)}(1) & \dots & T_K^{(L)}(1) & \dots & T_K^{(1)}(N-K) & \dots & T_K^{(L)}(N-K)\\ | & | & | & | & | & | & | & | & | & | \end{bmatrix}}_{\rm all \ accessible \ trajectories}
\end{equation}
of size $Kd^2\times L(N-K+1)$, which has the form of a Hankel matrix. Here, trajectories chunks can be thought of as vectors of length $Kd^2$. In the noiseless case, it is enough to take the linear span of the Hankel matrix columns to determine the subspace ${\cal C}_K$, but we need to process them further in order to distinguish the signal from background noise.

{\it Step 2.} One performs a singular value decomposition (SVD) of the Hankel matrix $H = USV^\dagger$, where $U$ and $V$ are isometric matrices and $S$ is the diagonal matrix with singular values on the diagonal arranged in non-increasing order.

{\it Step 3.} Following the results of \cite{gavish2014optimal} one separates ``noisy'' singular values from ``signal'' singular values by comparing them with a threshold
\begin{equation}
    s = \sigma\sqrt{2L(N-K+1)}f\left(\frac{Kd^2}{L(N-K+1)}\right),
\end{equation}
where $f$ is defined as follows $f(\beta) = \sqrt{2(\beta+1)+\frac{8\beta}{(\beta+1)+\sqrt{\beta^2+14\beta+1}}}$, $\sigma$ is the standard deviation of noise. If a singular value is greater (less) than the threshold, one considers it as a signal (noise) singular value. Since the quantum state tomography requires an enormous number of measurement outcomes to build a precise estimations of system's states, the size of measurements statistic is the main bottleneck of a tomographic experiment. Therefore, the observed noise is mainly the result of a statistical error caused by the finiteness of the quantum measurements statistics $N_{\rm samples}$ per time step. This allows one to estimate the standard deviation of noise $\sigma$ as $N_{\rm samples}^{-1/2}$. More precisely, the standard deviation $\sigma$ of the resulting noise reads $\sigma \approx N_{\rm samples}^{-1/2}$ for the standard tomographic protocols or $\sigma \approx N_{\rm samples}^{-1}$ for adaptive tomographic protocols with a protocol-specific prefactor~\cite{mahler2013adaptive}. If trajectories result from a Monte-Carlo simulation \cite{breuer2002theory} one can estimate $\sigma$ from the number of samples as $\sigma\approx N^{-1/2}_{\rm samples}$. The threshold $s$ differs from the threshold in \cite{gavish2014optimal} by a factor $\sqrt{2}$. This is motivated by the fact that the amplitude of the absolute value of a complex valued noise is $\sqrt{2}$ times larger. The intuition behind such a separation of singular values is simple: one uses the random matrix theory to determine the maximal singular value $s_{\rm max}$ of a random matrix with the standard deviation of a matrix elements $\sigma$. Then, $s \approx s_{\rm max}$ can be seen as a threshold that separates ``noisy'' singular values from ``signal'' singular values;

{\it Step 4.} One truncates the SVD of the Hankel matrix as follows
    \begin{equation}
        \hat{U} = U[:, :\eta],\quad
        \hat{V} = V[:, :\eta],\quad
        \hat{S} = S[:\eta, :\eta],
    \end{equation}
    where we use MATLAB/NumPy notations to represent slices of matrices (see Appendix~\ref{appendix:slicing}), $\eta$ is the number of singular values that are greater than $s$. The corresponding ``truncated'' Hankel matrix reads $\hat{H} = \hat{U}\hat{S}\hat{V}^\dagger$.

This algorithm provides a data-driven estimation of the genuine trajectories subspace ${\cal C}_K$, its dimension, and the dimension of the effective environment:
\begin{equation}
    {\cal C}_K \approx {\rm span}(\hat{U}),\quad 
    r\approx \eta,\quad 
    d_\rE^{\rm eff}\approx \ceil*{\sqrt{\eta / d^2}},
\end{equation}
where ${\rm span}(\cdot)$ denotes the linear span of matrix columns and $\eta$ serves as a data-driven estimation of the genuine trajectories subspace dimension $r$. But these are not the only useful outcomes of the algorithm: it also reduces noise in the measured trajectories.
\begin{figure}[h!]
    \centering
    \includegraphics{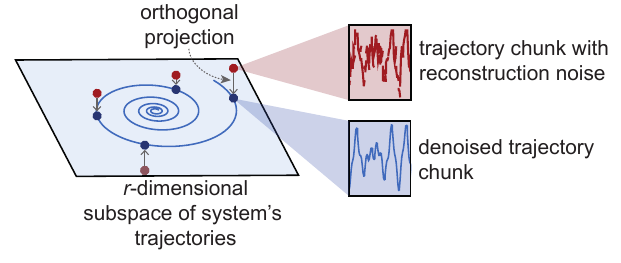}
    \caption{Illustrative interpretation of the PCA with automatic rank selection algorithm applied to quantum trajectories. The algorithm estimates the low-dimensional subspace of system's trajectories from measured trajectories and then projects trajectories chunks on this subspace. The orthogonal projection on this subspace eliminates the orthogonal to the subspace component of noise that leads to the noise reduction effect.}
    \label{fig:denoising illustration}
\end{figure}
The idea is that since noise does not preserve the subspace ${\cal C}_K$, it takes trajectories out of this subspace (see Fig. \ref{fig:denoising illustration}). Meanwhile, the orthogonal projection of trajectories back onto the subspace ${\cal C}_K$ reduces the effect of noise because the component of noise orthogonal to the subspace ${\cal C}_K$ is thus eliminated. Note, that one can reconstruct the whole trajectory of length $N$ from its chunks as shown in Fig.~\ref{fig:trajectory_slicing}{\bf b}. In order to establish a connection between the proposed denoising method with previously developed techniques for denoising and quantum tomographic reconstruction from limited data, we note that usually their main idea is to employ an assumption about underlying states model imposing certain restrictions on admissible states. In other words, these methods use prior information about an unknown state in addition to the observed measurements data, which mitigates requirements to the amount of data. For example, one can use the low-rank assumption to reconstruct a fairly pure state from limited data \cite{gross2010quantum}, or assume limited entanglement to make the reconstruction of a many-body state from measurement outcomes possible \cite{cramer2010efficient}. However, our technique utilizes another type of prior information. It uses prior information about the temporal structure of a system's dynamics, not about its state. It fixes some level of complexity of the dynamics that reduces the required amount of data and leads to noise reduction. Our denoising technique is compatible with  conventional methods of noise reduction, and in combination, they can lead to further improvement of data efficiency. We verify the performance of denoising in the numerical experiments discussed further below.

\subsection{Identification of the Markovian master equation governing trajectories dynamics}
Denoised trajectories $\hat{T}_K^{(i)}(t)$ which are columns of $\hat{H}$ can be also used to estimate $M$, which could be thought of as a predictive model from the machine learning point of view. The estimation can be written as the solution of the following optimization problem
\begin{eqnarray}
    \label{eq:central_optimization_problem}
    &&\underset{M}{\rm minimize \quad } \sum_{i=1}^L\sum_{t=0}^{N-K}\left\|\hat{T}_K^{(i)}(t+1) - M\left[\hat{T}_K^{(i)}(t)\right]\right\|_2^2,\nonumber\\
    &&{\rm subject \ to \quad } {\rm rank}\left(M\right) = \eta,
\end{eqnarray}
where $\|\cdot\|_2$ stands for the $2$-norm. The first line corresponds to the minimization of the difference between left and right hand sides of the equation $\hat{T}_K(t+1) = M[\hat{T}_K(t)]$ driving the dynamics of trajectories. The second line restricts consideration only to matrices of rank $\eta$. As the dynamics of trajectories is embedded in the subspace of dimension $\eta$, there is no need to consider matrices of other ranks. Note that $\eta$ is not only an estimate of $r$, but also the optimal choice of one of the hyperparameters of the predictive model. Another hyperparameter of the predictive model is the memory depth $K$. Search of $\eta$ corresponds to the selection of the predictive model. The larger the value of the hyperparameter is, the more expressible the corresponding model is (i.e., more complex data can be described), but the more data is needed to reconstruct $M$ (lower data efficiency). This is the bias–variance trade-off typical for the model selection tasks in machine leaning \cite{mehta2019high}. Usually one selects the best model by minimization of the error on a validation set, which is not used for model training. This model selection technique is called cross-validation \cite{arlot2010survey}, and requires splitting data on training and validation sets and exhaustive validation set's error minimization via brute force adjustment of hyperparameters. However, for some models, e.g. relevance vector machine \cite{tipping2001sparse, bishop2006pattern}, one selects the best model automatically without a validation set by using the Bayesian interpretation of the model selection task. The idea of such an automatic model selection follows the Occam's razor principle: one should select the simplest model among all possible that still yet manages to describe the observed data \cite{bishop2006pattern}. The correct model selection guarantees the best data efficiency and robustness to noise and corresponds to the best machine learning practices. One sees that a PCA algorithm with automatic determination of the subspace dimension solves such a model selection task. It chooses the model with the smallest value of the hyperparameter (the simplest model) that still describes the measured data. Therefore, our approach is equipped with the automatic model selection, which improves the data efficiency compared to previously proposed methods such as TTM \cite{cerrillo2014non} as discussed in Sec.~\ref{sec:comparison}. The effect of noise reduction also can be seen as a consequence of the correct model selection. It also makes the model less sensitive to selection of another hyperparameter $K$. One can safely overestimate $K$ by setting it knowingly large, the automatic model selection prevents the effect of overfitting. This makes the selection procedure for $K$ very simple: one can set $K$ to be comparable with $N$, e.g. $K = N / 2$. The only important restriction is not to make the number of trajectories $(N - K + 1)L$ in a data set too small. The proper selection of $\eta$ prevents overfitting. We show this numerically in Section~\ref{seq:Probing non-Markovian quantum dynamics with a predefined effective environment}.

The optimization problem Eq.~\eqref{eq:central_optimization_problem} is exactly solvable. Indeed, this is an example of linear regression among the most robust methods of machine learning. To write down the solution of Eq.~\eqref{eq:central_optimization_problem}, let us introduce two matrices
\begin{eqnarray}
    &&\hat{X} = \hat{H}[:, :-L] =  \begin{bmatrix} | & | & | & | & | & | & | & | & | & |\\ \hat{T}_K^{(1)}(0) & \dots & \hat{T}_K^{(L)}(0) & \hat{T}_K^{(1)}(1) & \dots & \hat{T}_K^{(L)}(1) & \dots & \hat{T}_K^{(1)}(N-K-1) & \dots & \hat{T}_K^{(L)}(N-K-1)\\ | & | & | & | & | & | & | & | & | & | \end{bmatrix},\nonumber\\
    &&\hat{Y} = \hat{H}[:, L:] =  \begin{bmatrix} | & | & | & | & | & | & | & | & | & |\\ \hat{T}_K^{(1)}(1) & \dots & \hat{T}_K^{(L)}(1) & \hat{T}_K^{(1)}(2) & \dots & \hat{T}_K^{(L)}(2) & \dots & \hat{T}_K^{(1)}(N-K) & \dots & \hat{T}_K^{(L)}(N-K)\\ | & | & | & | & | & | & | & | & | & | \end{bmatrix}.
\end{eqnarray}
The matrix $\hat{X}$ is the matrix $\hat{H}$ without $L$ last columns and the matrix $\hat{Y}$ is the matrix $\hat{H}$ without $L$ first columns. In other words, columns of $\hat{Y}$ are trajectories subsequent in time to the corresponding columns of $\hat{X}$. Using this fact, one can rewrite the optimization problem Eq.~\eqref{eq:central_optimization_problem} in the following form
\begin{eqnarray}
    &&\underset{M}{\rm minimize \quad }\left\|\hat{Y} - M\hat{X}\right\|_F^2,\nonumber\\
    &&{\rm subject \ to \quad } {\rm rank}\left(M\right) = \eta,
\end{eqnarray}
where $\|\cdot\|_F$ stands for the Frobenius norm, and $M$ is seen as a matrix rather than as an abstract linear map. The solution of this optimization problem thus reads:
\begin{equation}
    \label{eq:solution_of_the_central_optimization_problem}
    M = \hat{Y}\hat{X}^+,
\end{equation}
where $\hat{X}^+$ is the Moore–Penrose inverse of $X$. Indeed, $M$ has the rank $\eta$ by construction and it is the global minimum of the objective function. Note that it is more efficient both for numerical complexity and memory to find the solution Eq.~\eqref{eq:solution_of_the_central_optimization_problem} by using the \emph{dynamic mode decomposition} (DMD) method \cite{schmid2010dynamic,tu2013dynamic}, rather than straightforwardly calculate $YX^+$. In our numerical experiments we use the DMD to find $M$.

\begin{figure}[h]
    \centering
    \includegraphics[scale=0.7]{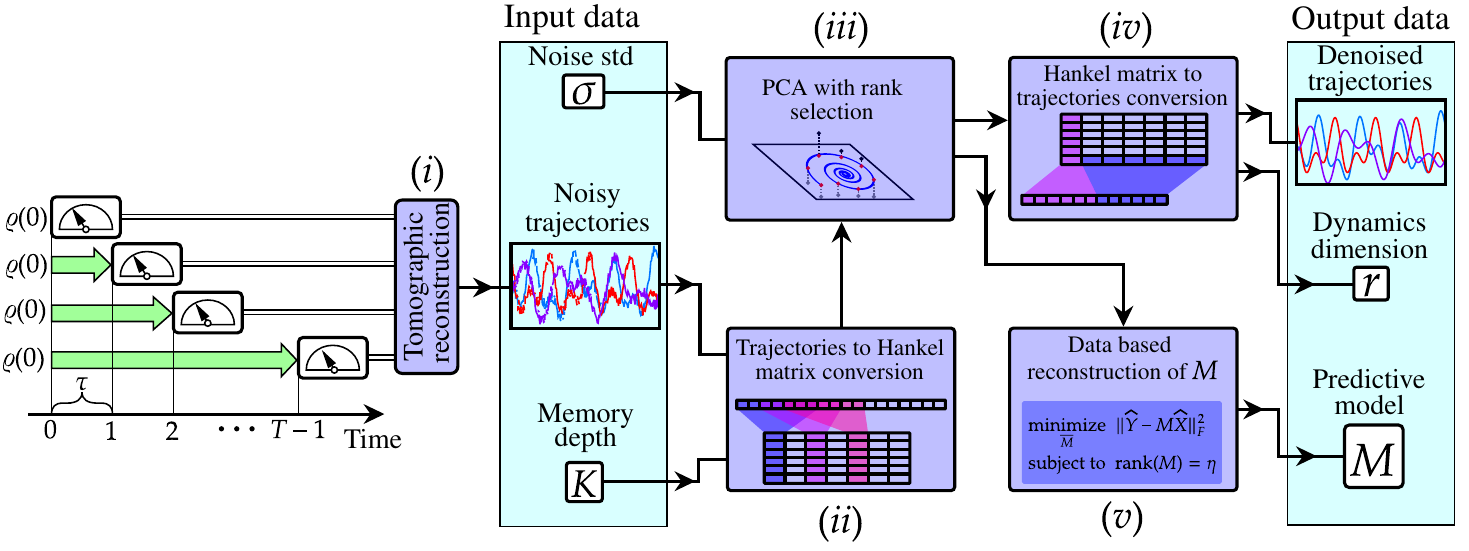}
    \caption{Summary of the proposed data processing scheme. (\emph{i}) One starts by collecting a data set of noisy quantum trajectories. Usually, it is obtained via tomographic reconstruction or numerical simulation. As input information the scheme also requires an estimation of the standard deviation of noise and an estimation of the sufficient $K$. The standard deviation can be estimated from the size of the quantum measurements statistics used per time step. An estimation of the sufficient $K$ does not requires precise tuning. It can be set knowingly large. The automatic model selection prevents overfitting for an overestimated $K$. One of the possible practical recipes of choosing $K$ consists in setting the maximal possible $K = N-1$ and iterative retraining the model while decrementing $K$, until the value of $\eta$ saturates. (\emph{ii}) In the next step, one slices trajectories into chunks of size $K$ as shown in Fig.~\ref{fig:trajectory_slicing}{\bf a}. (\emph{iii}) Then, one identifies the subspace of trajectories via a PCA with the automatic dimension selection. It not only reconstructs the space ${\cal C}_K$, but also allows to estimate the effective dimension of the environment $d_\rE^{\rm eff}$, and performs the automatic model selection and reduces noise in data. (\emph{iv}) Denoised chunks of trajectories are combined into a denoised version of initial whole trajectories of length $N$ as shown in Fig.~\ref{fig:trajectory_slicing}{\bf b}. (\emph{v}) Finally, one reconstructs $M$ from the denoised trajectories chunks by solving the optimization problem Eq.~\eqref{eq:central_optimization_problem}. The reconstructed $M$ allows predicting the dynamics of a non-Markovian quantum system and its eigenvalues are also eigenvalues of $\Phi$ which drives the joint system and environment dynamics.}
    \label{fig:central_pic}
\end{figure}

Finally, we summarize the data processing scheme in Fig.~\ref{fig:central_pic}, that basically includes two steps, ({\it i}) environment dimension identification described in the previous subsection and ({\it ii}) predictive model reconstruction described in the given subsection.
The proposed data processing scheme allows one to obtain (i) information on the quantum environment such as the dimension of the effective environment, eigenvalues of the quantum channel driving the discrete-in-time joint dynamics of the system and its environment, (ii) the master equation that predicts the system's dynamics, and (iii) denoised data set of trajectories. 
As inherent features of the scheme we note that all the used algorithms (i.e. PCA with automatic rank selection and the linear-regression-based reconstruction of $M$) converge with mathematical guarantees to the optimal solution and the scheme is equipped with the automatic model selection that improves data efficiency and robustness to noise.

\section{Numerical results}
\subsection{Probing non-Markovian quantum dynamics with a predefined effective environment}
\label{seq:Probing non-Markovian quantum dynamics with a predefined effective environment}
As the first example of non-Markovian quantum dynamics that we analyze with the proposed method, we consider a model with a predefined effective environment, i.e. with a predefined $\Phi^{\rm (rr)}$ driving the dynamics of the system and the effective environment as a whole. We choose $\Phi^{\rm (rr)}$ in such a way, that its dimension can not be reduced further.
This model is chosen to validate the claimed ability of the proposed method to reconstruct properties of the effective environment i.e. the effective dimension of the environment and eigenvalues of the joint system and environment dynamics. It can be easily done since all the properties of $\Phi^{\rm (rr)}$ are predefined and known. In this subsection we also (\emph{i}) validate the ability of our method to predict a non-Markovian quantum dynamics via the reconstructed $M$; (\emph{ii}) validate the ability of our method to reduce noise in data by checking how the projection of the observed trajectories on the principle subspace reduces the amplitude of noise; (\emph{iii}) study the sensitivity of the proposed method to a choice of the hyperparameter $K$; and finally (\emph{iv}) we study how the performance of the method depends on the size of data set, i.e. number of trajectories $L$ in a data set.

The discrete-in-time dynamics of the system reads
\begin{eqnarray}
    \label{eq:model_with_predefined_effective_environment}
    && \Phi^{\rm (rr)} = \exp\left(\tau {\cal L}\right),\nonumber\\
    &&\varrho^{\rm (r)}_\rSE(t+1) = \Phi^{\rm (rr)}\left[\varrho^{
    \rm(r)}_\rSE(t)\right],\nonumber\\
    && \varrho(t+1) = {\rm Tr}_\rE\left(\varrho^{\rm(r)}_\rSE(t+1)\right),
\end{eqnarray}
where $\varrho^{\rm(r)}_\rSE\in\mathbb{C}^{dd^{\rm eff}_\rE\times dd^{\rm eff}_\rE}$ is the joint system and effective environment density matrix that also plays the role of the state in the relevant subspace, $d_\rE^{\rm eff}$ is the dimension of the effective environment, $\varrho$ is the state of the system, ${\cal L}$ is the Gorini–Kossakowski–Sudarshan–Lindblad (GKSL) generator \cite{gorini1976completely, lindblad1976generators, accardi2013quantum, breuer2002theory, alicki2007quantum} of the Markovian quantum dynamics of the system and the effective environment, $\tau$ is the constant discrete time step. One generates the GKSL generator ${\cal L}$ randomly as described in the Appendix~\ref{appendix:random_linbladians_gen}. The randomness of ${\cal L}$ prevents further dimensionality reduction of the effective environment.

We use the model above to generate a number of data sets. We fix $d=2$ (the system is a qubit) and for $d_\rE^{\rm eff}$ ranging from $2$ to $6$ generate a set of random GKSL generators. For each GKSL generator and for total simulation times $N=150, \ \text{and} \ 200$ we simulate $L=4$ quantum trajectories $\left\{T^{(i)}_N(0)\right\}_{i=1}^L$ with time step $\tau = 0.2$ using Eq.~\eqref{eq:model_with_predefined_effective_environment}. As an initial joint state for each trajectory we take $\varrho_\rSE(0) = \ket{\psi}\bra{\psi}\otimes \varrho_\rE(0)$, where $\ket{\psi}$ is sampled uniformly from the Bloch sphere, $\varrho_\rE(0)={\rm Tr}_\rS \varrho^{\rm st}_\rSE$, and $\varrho^{\rm st}_\rSE$ is the stationary state of ${\cal L}$, i.e. ${\cal L}\left[ \varrho^{\rm st}_\rSE\right]=0$. It means that the system and the environment are in a joint thermalized state before the observation process, and at the moment the process starts, we replace the system by a new random pure state. To address dynamics prediction accuracy, for each value of $d_\rE^{\rm eff}$, we generate an additional `test' trajectory $T^{\rm test}_N(0)$, which is not employed in the training procedure, i.e. its initial state is also randomly generated but different with initial states of $\{T^{(i)}_N(0)\}_{i=1}^L$. In order to simulate noise appearing during data acquisition of trajectories, we add i.i.d. Gaussian noise with zero mean and variance $\sigma^2$ to the real and imaginary parts of each entry of all trajectories (both employed in the training and the test trajectory). In what follows, we consider $\sigma$ to be known in advance, because in a typical tomographic experiment or in the Monte-Carlo simulation one can estimate $\sigma$ from the statistics size. We also note that the considered values of $\sigma$ are consistent with typical numbers of samples processed in tomographic experiments $N_{\rm samples} \sim 10^{3} \ldots 10^6$ (see e.g.~\cite{mahler2013adaptive}). To demonstrate the denoising ability of our approach we keep record of both noisy and noiseless versions of each trajectory, i.e. each trajectory of all data sets $\{T^{(i)}_N(0)\}_{i=1}^L$ and all test trajectories $T^{\rm test}_N(0)$.

\begin{figure}[ht]
    \centering
    \includegraphics[width=0.99\linewidth]{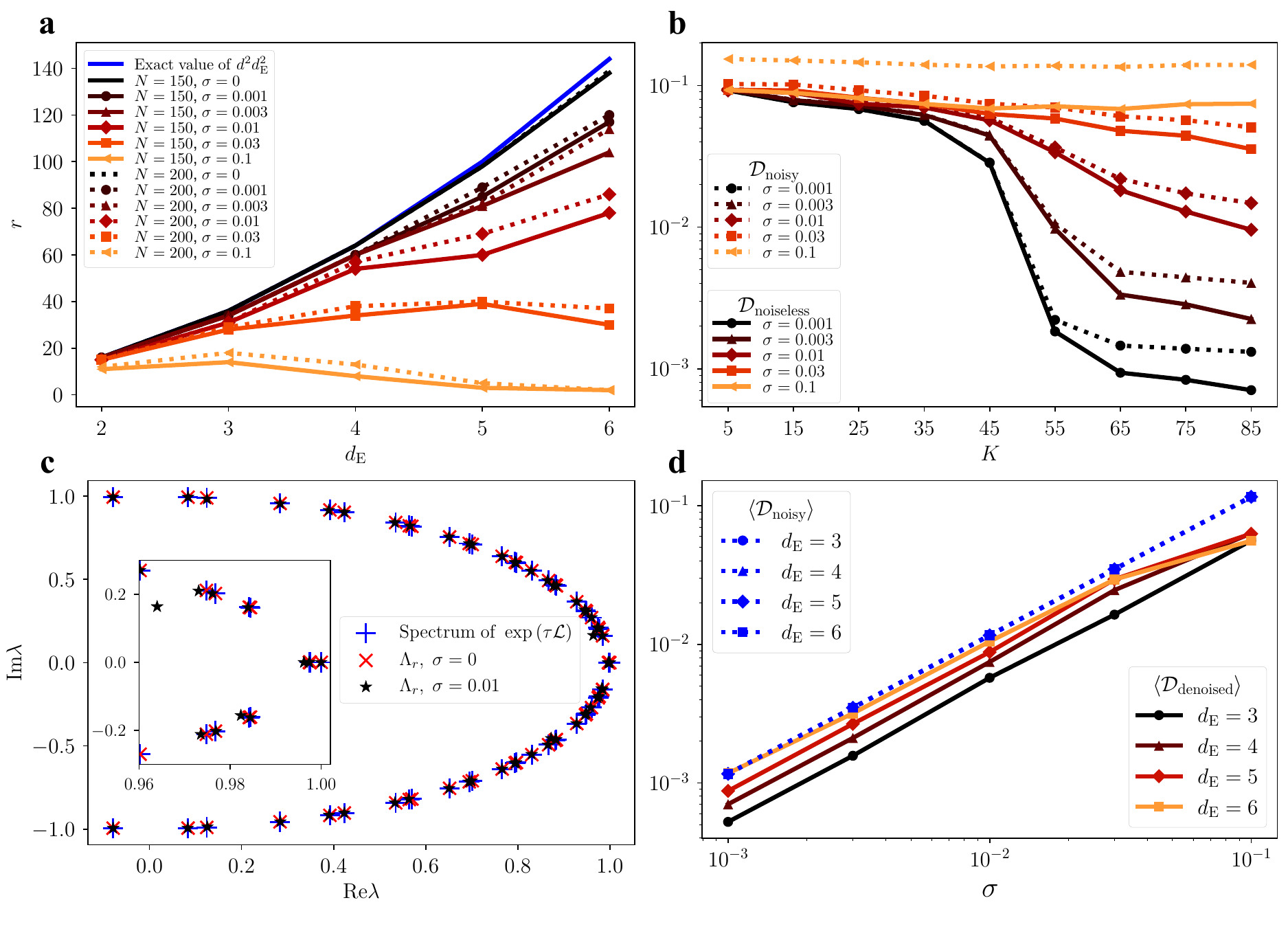}
    \caption{Results obtained by the application of the proposed method to the non-Markovian dynamics with a predefined effective environment (Eq.~\eqref{eq:model_with_predefined_effective_environment}). {\bf a} A comparison of the exact dimension of the trajectories subspace $d^2[d_\rE^{\rm eff}]^2$ with the data driven reconstruction $r$ of it for different data set trajectories lengths $N$ and noise levels $\sigma$.
    {\bf b} The distance between the prediction and the noiseless test trajectory ${\cal D}_{\rm noiseless}$ and the distance between the prediction and the noisy test trajectory ${\cal D}_{\rm noisy}$ as functions of memory depth $K$ for different levels of noise $\sigma$.
    {\bf c} Comparison of the spectrum of the quantum channel $\exp(\tau{\cal L})$ driving the dynamics of the joint system and effective environment density matrix with the eigenvalues of $M$ reconstructed from data for $d_\rE=4$.
    {\bf d} Comparison of the distance between noisy and noiseless data sets $\left\langle{\cal D}_{\rm noisy}\right\rangle$ and the distance between denoised and noiseless data sets $\left\langle{\cal D}_{\rm denoised}\right\rangle$ for several different $\sigma$ and $d_\rE$.}
    \label{fig:first_part_of_results}
\end{figure}

We start discussion of the results of data processing from the reconstruction of the trajectories subspace dimension $r$ and the effective dimension of the environment $d^{\rm eff}_\rE$.
The comparison of the exact $r$ and the reconstructed from data one for the fixed memory depth hyperparameter $K=75$, several noise amplitudes $\sigma$, and two values of a data set trajectories length $N$, is presented in Fig.~\ref{fig:first_part_of_results}{\bf a}. One can see that the reconstructed $r$ underestimates the exact value of $r$ that is equal to $d^2d_\rE^2$ though approaches it with decreasing the noise level and increasing the length of trajectories forming a data set. In the case of $\sigma=0$, reconstructed $r$ precisely fits the exact one. The tiny difference appearing for large $r$ is due to negligible memory effects that do not fit into given memory length $K$. For finite $\sigma$ it is impossible to recognize some weak memory effects in front of noise, this leads to underestimated $r$ value. The situation becomes critical when $\sigma=0.1$, as the algorithm is almost unable to recognize the signal against the background noise. We also compare the exact effective dimensions of the environment $d_\rE^{\rm eff}$ with its data driven reconstruction $\ceil*{\sqrt{r / d^2}}$, where $r$ is reconstructed from the data. The comparison is given in Table~\ref{tab:effective-dims}. We see that for small noise amplitudes and sufficiently long trajectories in a data set, the value of $d_\rE^{\rm eff}$ is reconstructed exactly.

\begin{table}[]
    \centering
    \begin{tabular}{c|c|c|c|c|c|c|}
\cline{2-7}
\multicolumn{1}{c|}{}                     & \multicolumn{2}{c|}{$\sigma=10^{-1}$} & \multicolumn{2}{c|}{$\sigma=10^{-2}$} & \multicolumn{2}{c|}{$\sigma=10^{-3}$} \\ \cline{2-7} 
\multicolumn{1}{c|}{}                     & $N=150$           & $N=200$           & $N=150$           & $N=200$           & $N=150$           & $N=200$           \\ \hline
\multicolumn{1}{|c|}{$d_\rE^{\rm eff}=2$} & \textbf{2}                 & \textbf{2}                 & \textbf{2}                 & \textbf{2}                 & \textbf{2}                 & \textbf{2}                 \\ \hline
\multicolumn{1}{|l|}{$d_\rE^{\rm eff}=3$} & 2                 & 2                 & \textbf{3}                 & \textbf{3}                 & \textbf{3}                 & \textbf{3}                 \\ \hline
\multicolumn{1}{|l|}{$d_\rE^{\rm eff}=4$} & 2                 & 2                 & \textbf{4}                 & \textbf{4}                 & \textbf{4}                 & \textbf{4}                 \\ \hline
\multicolumn{1}{|l|}{$d_\rE^{\rm eff}=5$} & 2                 & 2                 & 4                 & \textbf{5}                 & \textbf{5}                 & \textbf{5}                 \\ \hline
\multicolumn{1}{|l|}{$d_\rE^{\rm eff}=6$} & 1                 & 1                 & 5                 & 5                 & 5                 & \textbf{6}                 \\ \hline
\end{tabular}
    \caption{Comparison of the exact effective dimensions of the environment $d_\rE^{\rm eff}$ with reconstructed  from data ones for different amplitudes of noise $\sigma$ and different trajectories lengths $N$. Cases where the data driven estimation is exact are shown in bold font.}
    \label{tab:effective-dims}
\end{table}

\begin{figure}[h!]
    \centering
    \includegraphics[scale=0.25]{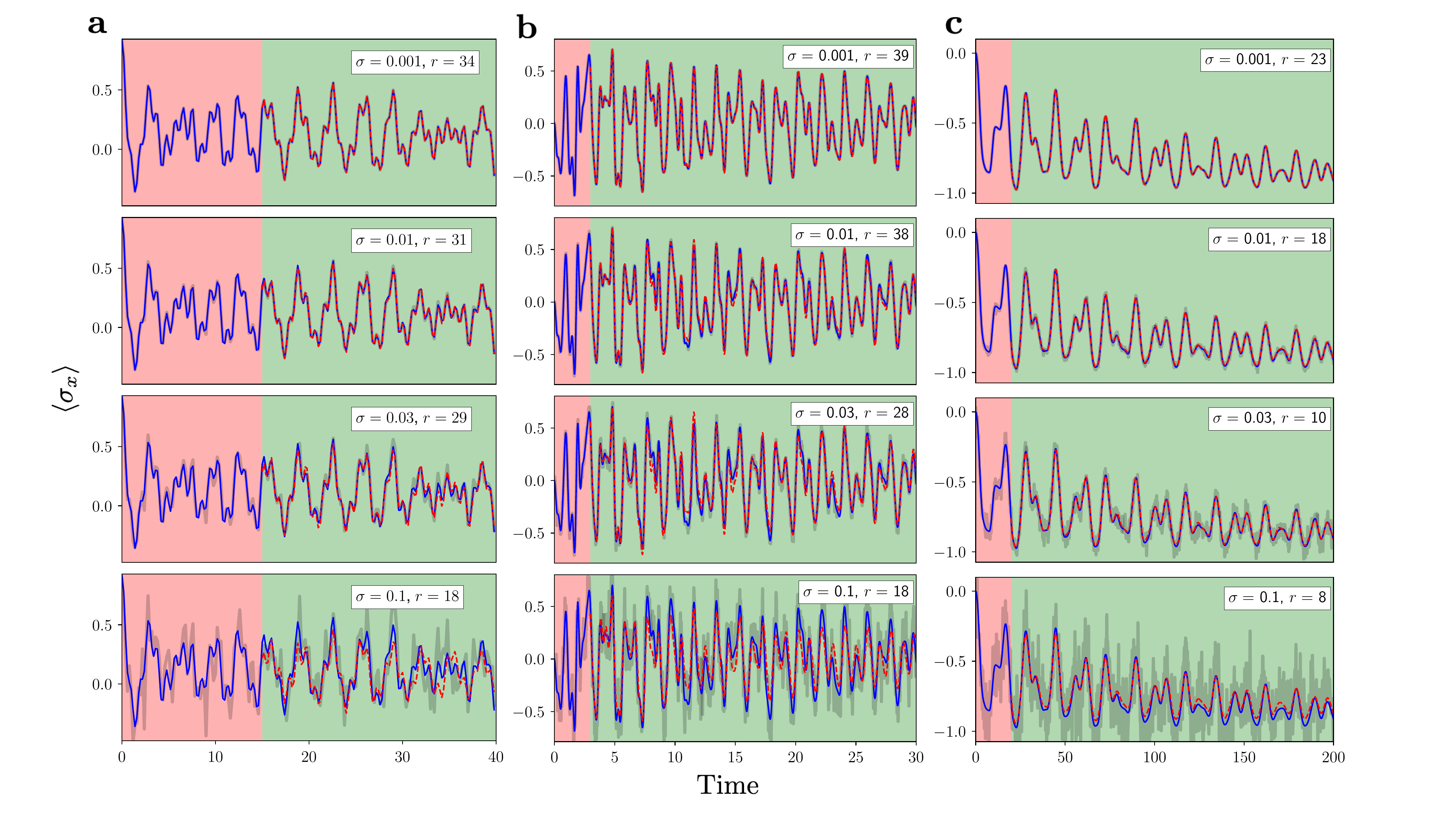}
    \caption{Comparison of the noiseless dynamics of $\langle\sigma_x\rangle = {\rm Tr}\left(\sigma_x\varrho\right)$ computed for the test trajectory (blue curve), noisy version of this dynamics (gray thick curve) and the data based prediction (red dashed curve) within \textbf{a} the model with the predefined effective environment (Eq.~\eqref{eq:model_with_predefined_effective_environment}) of dimension $d_\rE = 3$, \textbf{b} The damped Jaynes-Cummings model (Eq.~\eqref{eq:jc_lindblad}) with parameters $\gamma=0.05$, $g=2.5$, $\alpha=1.1$ and \textbf{c} the spin-boson model (Eq.~\eqref{eq:sb_hamiltonian}) with parameters $\gamma=0.05$, $g=0.5$, $\Delta=0.5$. The value of $r$ shows the estimated from data dimension of trajectories subspace. The region where the noisy trajectory chunk $T^{\rm test}_K(0)$ is assumed to be known and employed for prediction of the further dynamics is highlighted by the red color. The region where one builds the prediction of the further dynamics $T^{\rm test}_{N-K}(K)$ is highlighted by the green color.}
    \label{fig:dynamics_prediction}
\end{figure}

Next, we validate the ability of the proposed method to predict the dynamics of the system. The prediction is built as follows. One reconstructs $M$ that describes Markovian dynamics of trajectories from a noisy data set. Then one uses $M$ to predict dynamics of a test trajectory. One takes a noisy chunk $T_K(0)$ of the test trajectory, also seen as an initial state of Eq.~\eqref{eq:MarkovianM} or as a ``prehistory'' of non-Markovian dynamics, and propagates it forward in time by using the reconstructed from data $M$ and Eq.~\eqref{eq:MarkovianM}. In Fig.~\ref{fig:dynamics_prediction}{\bf a} we demonstrate the agreement between the predicted behavior of the test trajectory and the test trajectory itself for the following parameters of the data set used for the reconstruction of $M$: $N=200$, $d_\rE=3$ and several different values of $\sigma$. The value of $K$ is set to $75$. One sees that the proposed method predicts non-Markovian dynamics of the system correctly. Even if a data set and a test trajectory are effected by noise with reasonably high standard deviation $\sigma=0.1$, the approach is capable to predict the system's dynamics at a small time scale.

Then we compare eigenvalues of $M$ with eigenvalues of the quantum channel $\Phi^{\rm (rr)} = \exp\left(\tau{\cal L}\right)$ driving the joint system and effective environment dynamics. The results for $M$ reconstructed from a data set with $d_\rE=4$, $K=75$, $N=200$ in the case of noise absence ($\sigma=0$) and in the case with noise ($\sigma=0.01$) are presented in Fig.~\ref{fig:first_part_of_results}{\bf c}. One can see that there is a perfect coincidence between exact and reconstructed eigenvalues in the case without noise, which agrees well with the theory (see Section~\ref{seq:theory}), while in the case with noise the obtained eigenvalues are slightly shifted, and a tiny part of them only is lost. The latter fact can be explained by the indistinguishability of some eigenmodes dynamics from the noise. However, even in the presence of noise, one sees that the method provides valuable information such as the eigenvalues of the joint dynamics.

\begin{figure}[ht]
    \centering
    \includegraphics[scale=0.7]{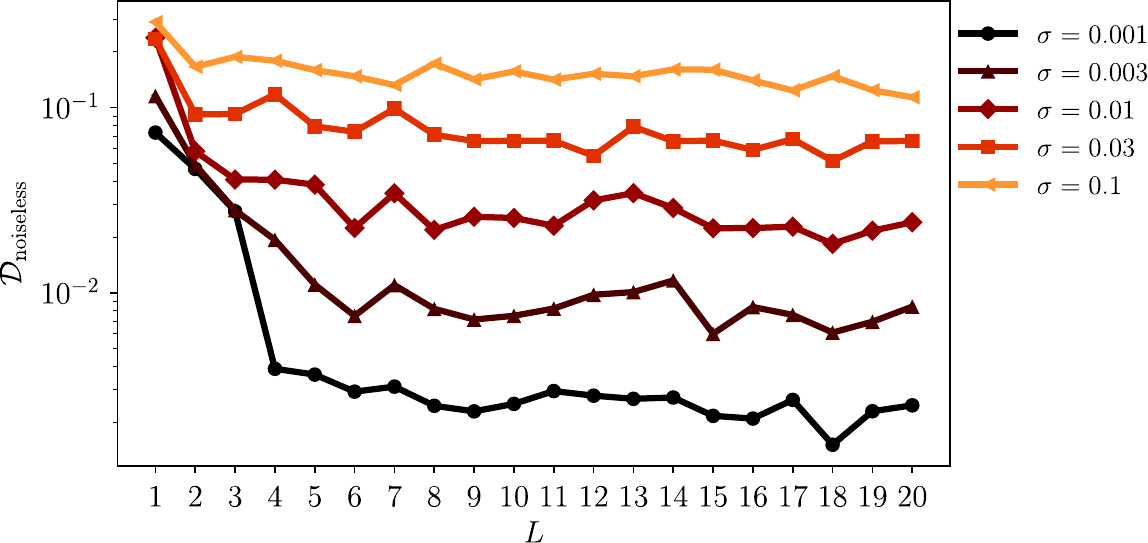}
    \caption{The distance between prediction and noiseless test trajectory ${\cal D}_{\rm noiseless}$ plotted against $L$ and $\sigma$ for the model with predefined effective environment (Eq.~\eqref{eq:model_with_predefined_effective_environment}). The value of $K$ is set equal to $75$.}
    \label{fig:D_vs_L}
\end{figure}

We also study how the selection of the memory depth $K$ that is a hyperparameter of the model and number of trajectories $L$ (data set size) impact the accuracy of the prediction. For this purpose we introduce the distance between two quantum trajectories $T^{(a)}_{R}(t)$ and $T^{(b)}_{R}(t)$ of length $R$ at time $t$ as follows
\begin{equation}
    \label{eq:dist_between_trajectories}
    {\cal D}\left(T^{(a)}_{R}(t), T^{(b)}_{R}(t)\right) = \frac{1}{R}\sum_{k=0}^{R-1}\left\|\varrho^{(a)}(t+k) - \varrho^{(b)}(t+k)\right\|_1,
\end{equation}
where states $\varrho^{(a)}$ and $\varrho^{(b)}$ are taken from the trajectories $T^{(a)}_R(t)$ and $T^{(b)}_R(t)$ correspondingly, and $\|\cdot\|_1$ denotes a trace norm \cite{chuang1997prescription}. Let ${\cal D}_{\rm noisy}$ be the distance Eq.~\ref{eq:dist_between_trajectories} between the prediction and the noisy version of a test trajectory $T_{N-K}^{\rm test}(K)$ and ${\cal D}_{\rm noiseless}$ be the distance Eq.~\ref{eq:dist_between_trajectories} between the prediction and the noiseless version of a test trajectory $T_{N-K}^{\rm test}(K)$. These distances show how close the predicted trajectory to the noisy test trajectory and to the noiseless test trajectory and thus can be used to validate the accuracy of the prediction. The behavior of the distance as a function of $K$ is presented in Fig.~\ref{fig:first_part_of_results}{\bf b}. As one can see, there is a `saturation' value of $K$, starting from which the accuracy of prediction is mainly determined by the noise level. This means that the accuracy of the prediction is not sensitive to the choice of $K$, i.e. there is no effect of overfitting when the accuracy becomes worse with an increase of the model complexity. One can set $K$ knowingly large and this does not lead to the effect of overfitting. This is the consequence of the proper automatic model selection. To validate how the accuracy of prediction depends on choice of $L$ we generated additional data sets for $L$ ranging from $1$ to $20$, $d^{\rm eff}_\rE = 4$, $N=200$ and for several different $\sigma$. For predictions based on these data sets, we plot the dependence of $D_{\rm noiseles}$ on $L$ and $\sigma$ in Fig.~\ref{fig:D_vs_L}. One can observe that the accuracy rapidly improves with growing $L$ at the beginning and then saturates. The saturation effect is caused by the fact that the prediction is based on an initial noisy prehistory of constant size (independent on $L$) and this noise can not be eliminated via an increase of data set size.

Finally, we validate the ability of the method to reduce noise in data. For this purpose we introduce the distance between data sets as the averaged over all trajectories distance introduced in Eq.~\ref{eq:dist_between_trajectories} and denote this distance as $\langle{\cal D}\rangle$. One needs to consider the distance between denoised and noiseless data sets $\langle{\cal D_{\rm denoised}}\rangle$ and the distance between noisy and noiseless data sets $\langle{\cal D_{\rm noisy}}\rangle$. If $\langle{\cal D_{\rm denoised}}\rangle$ is systematically smaller than $\langle{\cal D_{\rm noisy}}\rangle$ then the denoising procedure works correctly. We plot $\langle{\cal D_{\rm denoised}}\rangle$ and $\langle{\cal D_{\rm noisy}}\rangle$ for $T=200$, $K=75$ and several different values of $\sigma$ and $d_\rE$ in Fig.~\ref{fig:first_part_of_results}d. One can see that $\langle{\cal D_{\rm denoised}}\rangle$ is systematically smaller than $\langle{\cal D_{\rm noisy}}\rangle$. This fact supports our claim on the correctness of the denoising procedure. This completes the validation of the proposed method.

\subsection{Probing dynamics of dissipative Jaynes-Cummings model}
As another example of non-Markovian quantum dynamics, we consider the dynamics of a two-level atom (the system) interacting with a decaying bosonic mode (the environment) via the Jaynes-Cummings (JC) interaction~\cite{jaynes1963comparison, fink2008climbing}.
Note that in contrast to the previous example, the effective environment of a spin is a bosonic mode that is infinite-dimensional. However, in the previous example we considered a random GKSL generator driving the joint dynamics. Its randomness prevents further dimensionality reduction of the effective environment. In the given example, the bosonic mode can be truncated since we consider a case when the mode is not highly excited. This means that one can build an effective environment of finite dimension. The interesting question is whether the proposed algorithm can detect a low-dimensional effective environment from data. To address this question, we probe our method on this model and try to identify a low-dimensional effective environment. The details of the model are given below. The joint dynamics of the atom and the bosonic mode within the JC model is driven by the Lindblad equation that reads
\begin{eqnarray}
\label{eq:jc_lindblad}
    &&\frac{d\varrho_\rSE}{dt} = -i[H_{\rm JC}, \varrho_\rSE] + \gamma \left(a \varrho_\rSE a^\dagger - \frac{1}{2}a^\dagger a \varrho_\rSE - \frac{1}{2}\varrho_\rSE a^\dagger a\right), \nonumber \\
    &&H_{\rm JC} = a^\dagger a + \frac{1}{2}\sigma_z + \frac{g}{2}\left(\frac{\sigma_x + i\sigma_y}{2}a + {\rm h.c.}\right),
\end{eqnarray}
where $H_{\rm JC}$ is the JC model Hamiltonian, $a$($a^\dagger$) is the bosonic mode annihilation (creation) operator, $\sigma_x, \sigma_y, \sigma_z$ are the Pauli matrices, $g$ is the interaction strength, and $\gamma$ is the bosonic mode dissipation rate. Taking the partial trace w.r.t. the bosonic mode, one obtains the density matrix of the atom, $\varrho(t) = {\rm Tr}_\rE(\varrho_\rSE(t))$, that experiences non-Markovian dynamics. As in the previous example, for $\gamma=0.05$, $g=2.5$ and various amplitudes of noise $\sigma$ we prepare a number of noisy data sets $\left\{T^{(i)}_N(0)\right\}_{i=1}^L$ and noisy and noiseless versions of test trajectories $T^{\rm test}_N(0)$ for each data set using the model above. We simulate the joint atom and mode dynamics driven by Eq.~\eqref{eq:jc_lindblad} at successive time steps of length $\tau = 0.03$, and compute the corresponding trajectories of the system by taking a partial trace over the environment. The initial joint atom and mode state is taken in the factorized form
\begin{eqnarray}
    && \varrho_\rSE(0) = \ket{\psi}\bra{\psi}\otimes\ket{\alpha}\bra{\alpha},
\end{eqnarray}
where $\ket{\alpha}$ is the coherent state of the bosonic mode, and $\ket{\psi}$ is sampled uniformly from the Bloch sphere. For all data sets we fix $\alpha = 1.1$. To proceed with numerical simulation of Eq.~\eqref{eq:jc_lindblad} we truncate the infinite-dimensional Hilbert space of the bosonic mode keeping such a number of eigenstates with the lowest energy that guarantees conservation of $>95\%$ of the initial environment state $\varrho_\rE(0)$ probability mass. The truncation is needed only for the tractability of numerical calculations necessary for data set generation. The limit of applicability of our data processing method is not sensitive to the genuine dimensionality of the environment; it can be either finite or infinite-dimensional. As mentioned before, only the finiteness of the memory matters. We fix $L = 2$, $K = 100$ and $N = 1000$ for all data sets.\\
We apply our method to generated data sets and reconstruct $r$ and $M$. As in the previous example, we use $M$ to predict behavior of the test trajectory. The comparison of the prediction and the test trajectory for each amplitude of noise is shown in Fig.~\ref{fig:dynamics_prediction}{\bf b}. One can see that one predicts dynamics of the atom even in presence of relatively high noise. We also note, that in all considered cases $r \leq 39$, while the dimension $d^2\left[d_\rE^{\rm eff}\right]^2$ of the joint system and environment density matrix after truncation of the bosonic mode is 100, which is well above $39$. This justifies the truncation of the bosonic mode and shows the finiteness of $d_\rE^{\rm eff}$ even though the genuine environment is infinite. The main outcome of this numerical experiment is that the proposed method is able to identify a finite-dimensional effective environment even in the case of infinite-dimensional genuine environment.

\subsection{Probing the spin-boson model dynamics}
The third example of non-Markovian quantum dynamics, which we analyze with our method, is the dynamics of a two-level atom coupled with a set of non-interacting bosonic modes. We refer to this model as the spin-boson model~\cite{de2017dynamics}. We consider the dynamics of a two-level atom in the limit of a continuum of bosonic modes. In this case there is no a straightforward way to extract a finite dimensional effective environment using first-principle numerical modeling. We test the ability of the proposed method to do this from a data set as well as the ability to predict the non-Markovian dynamics of the spin. We also study how the dimension  $r$ reconstructed from data depends on some parameters of the spin-boson model and show that $r$ serves as a non-Markovianity or complexity measure in this case.
\\
The Hamiltonian of the spin-boson model reads
\begin{equation}
    \label{eq:sb_hamiltonian}
    H_{\rm SB} = \frac{1}{2}\sigma_z + \frac{1}{2}\Delta \sigma_x + \sum_{k} \omega_k a^\dagger_k a_k + \sigma_z X,\quad X = \sum_k \frac{g_k}{\sqrt{2\omega_k}}\left(a^\dagger_k + a_k\right),
\end{equation}
where $\Delta$ is the tunneling matrix element, $\omega_k$ are frequencies of bosonic modes, $a_k(a^\dagger_k)$ are annihilation (creation) operators of bosonic modes, and $g_k$ is the strength of interaction between the atom and the $k$-th mode. As before, we consider the atom as a system and the set of bosonic modes as an environment. The atom affected by the bosonic modes experiences non-Markovian quantum dynamics that can be analyzed by our method. We consider the case when the initial state of bosonic modes is the ground state uncorrelated with the initial state of the atom. In this case, the influence of the environment of non-interacting bosonic modes on the system is fully described by the two-time correlation function of bosonic modes
\begin{equation}
    C(t) = \bra{\rm vac} e^{\imath \sum_k \omega_ka^\dagger_ka_kt}Xe^{-\imath \sum_k \omega_ka^\dagger_ka_kt}X\ket{\rm vac} = \int_0^\infty \frac{J(\omega)}{\pi}\exp\left(-\imath\omega t\right), \quad J(\omega) = \pi\sum_k \frac{g_k^2}{2\omega_k}\delta(\omega - \omega_k),
\end{equation}
where $\ket{\rm vac}$ is the ground state of bosonic modes, $J(\omega)$ is the spectral density.
We consider the limit of continuum bosonic modes when the gap between frequencies of neighboring modes $\omega_{k+1} - \omega_{k}$ vanishes. In this limit we choose the following spectral density:
\begin{equation}
    J(\omega) = \frac{\gamma g^2 \omega}{(\omega^2 - \omega_0)^2 + \gamma^2\omega^2},
\end{equation}
where $\omega_0$ is the resonance frequency, $\gamma$ is the width of the spectral function, $g$ is the aggregated interaction strength.
\\
We simulate the dynamics of the atom with an arbitrary pure initial state $\ket{\psi}$ using the numerical approach and code developed in \cite{lambert2019modelling}. As before, we prepared a number of data sets $\left\{T_N^{(i)}(0)\right\}_{i=0}^L$ and corresponding test trajectories $T_N^{\rm test}(0)$ specified by various amplitudes of noise and parameters of the spin-boson model. The initial state of the two-level atom $\ket{\psi}$ is sampled uniformly from the Bloch sphere for each trajectory of a data set. We fix $L=2$, $\tau=0.15$, $K=100$, and $N=1000$ for all data sets. As before, we reconstruct $M$ from a data set and use it to predict the behavior of the test trajectory. The comparison of the prediction and the test trajectory is given in Fig.~\ref{fig:dynamics_prediction}{\bf c}. One can see again, that even for the relatively high amplitude of noise, the approach is capable of predicting the atom dynamics.

\begin{figure}[h!]
    \centering
    \includegraphics[scale=0.4]{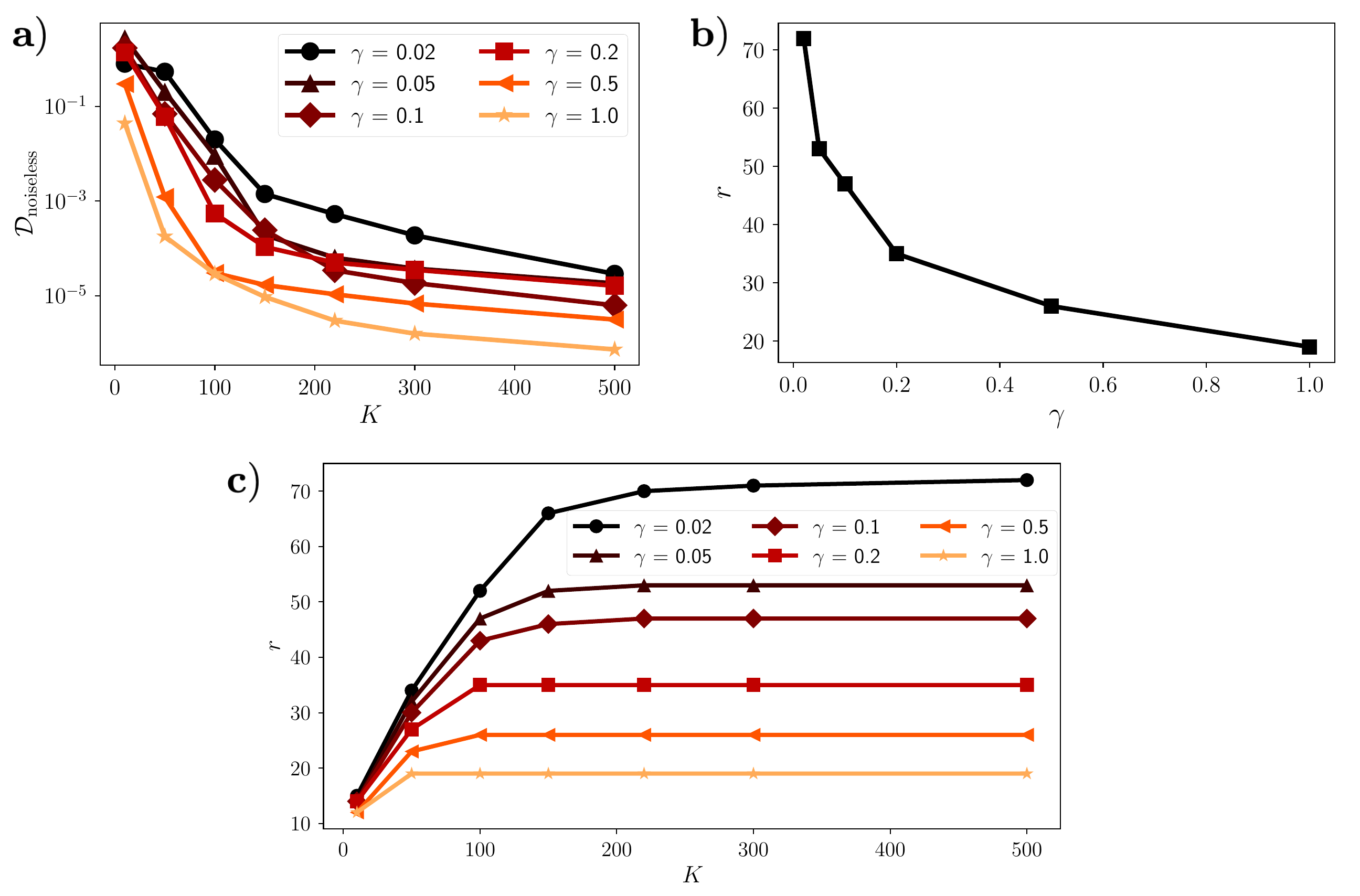}
    \caption{Analysis of the spin boson model (Eq.~\eqref{eq:sb_hamiltonian}) within the proposed method.
    {\bf a} Accuracy of prediction for different values of $K$ and $\gamma$. {\bf b} Trajectories subspace dimension for different values of $K$ and $\gamma$. {\bf c} Dependence of the trajectories subspace dimension on $\gamma$ for $K=500$.}
    \label{fig:sb_plot}
\end{figure}

In order to demonstrate on a concrete example that the trajectories space dimension $r$ can be also considered as a measure of non-Markovianity and system's dynamics complexity, we study a relation between $r$ and the value of the parameter $\gamma$ of the considered spin-boson model. Note that $\gamma$ determines the width of the two-time correlation function $C(t)$. The smaller $\gamma$ is the bigger the width of the correlation function is. Therefore the smaller $\gamma$ is the stronger memory effects are, and the more challenging the atom dynamics simulation is. We also consider different values of $K$ in order to understand how the memory depth of the dynamics depends on $\gamma$. We generate several noiseless data sets for $L = 4$, $T = 1000$, $\tau = 0.15$ and different values of $\gamma$. For each value of $\gamma$ we process the corresponding data set by our method. We consider different values of the hyperparameter $K$ and the fixed value of $\sigma=10^{-6}$ in order to cut off the numerical simulation error. We start from the analysis of the prediction accuracy. In Fig.~\ref{fig:sb_plot}a we show how the accuracy of prediction ${\cal D}_{\rm noiseless}$ depends on $K$ and $\gamma$. One can observe that a steep improvement of the accuracy then changes to a smooth saturation regime with increasing $K$. Note that for smaller $\gamma$ one has worse accuracy of prediction. This observation agrees well with the fact that for smaller $\gamma$, dynamics becomes more complicated. Then we turn to the analysis of $r$. In Fig.~\ref{fig:sb_plot}b we show how the minimal Markovian embedding dimension depends on different values of $K$ and $\gamma$. One can see that value of $r$ saturates starting from some sufficiently large value of $K$. Note also that for smaller $\gamma$ the value of $K$, for which $r$ saturates, is bigger. This supports the fact that for smaller $\gamma$ one has deeper memory. But the most important observation is that for smaller $\gamma$ one has $r$ bigger. In Fig.~\ref{fig:sb_plot}c, for $K=500$ (which is well above the memory depth for all considered $\gamma$) we show the dependence of $r$ on $\gamma$ explicitly. One sees that $r$ decreases with increasing $\gamma$, which is in accord with interpretation of $r$ as the complexity of a non-Markovian quantum dynamics or a measure of non-Markovianity.

\section{Comparison with existing data-driven methods of non-Markovian quantum dynamics identification}
\label{sec:comparison} 
In this section, we compare the given method of non-Markovian quantum dynamics identification with previously proposed ones, especially focusing attention on the related ``black-box'' methods namely the TTM \cite{cerrillo2014non, rosenbach2016efficient, buser2017initial, kananenka2016accurate, gelzinis2017applicability, chen2020non, gherardini2021transfer, pollock2018tomographically}, and a method based on a Recurrent Neural Network (RNN) parametrization of a predictive model \cite{banchi2018modelling}. We evaluate all the methods by means of the following metrics: ability of a method to extract information about the environment of a non-Markovian quantum system, data efficiency, and robustness to noise.

The first and foremost distinguishing feature of the proposed method is its ability to extract information about the quantum environment of a system by observing only quantum trajectories of a system. It estimates the effective dimension of the environment and reconstructs eigenvalues of the joint system and environment dynamics. The effective dimension of the environment shows the size of the environment fraction that mutually exchanges information with a system. The remaining fraction of the environment is seen as a decoupled part, since it only absorbs information and does not bring any information back. The eigenvalues of the joint system and environment dynamics are essentially spectroscopic data that defines the physics of the system and environment as a whole. All previously proposed methods do not reconstruct any information about the system's quantum environment, except maybe attempts to reconstruct the effective dimension of the environment by looking for the ``elbow'' in the validation curve plot \cite{luchnikov2020machine, guo2020tensor} while fitting the observed data by process-tensor-based methods. However, this method requires a separate validation data set and multiple retraining of a process-tensor-based model, while the proposed method is free from these drawbacks.

\begin{figure}[ht]
    \centering
    \includegraphics[scale=1]{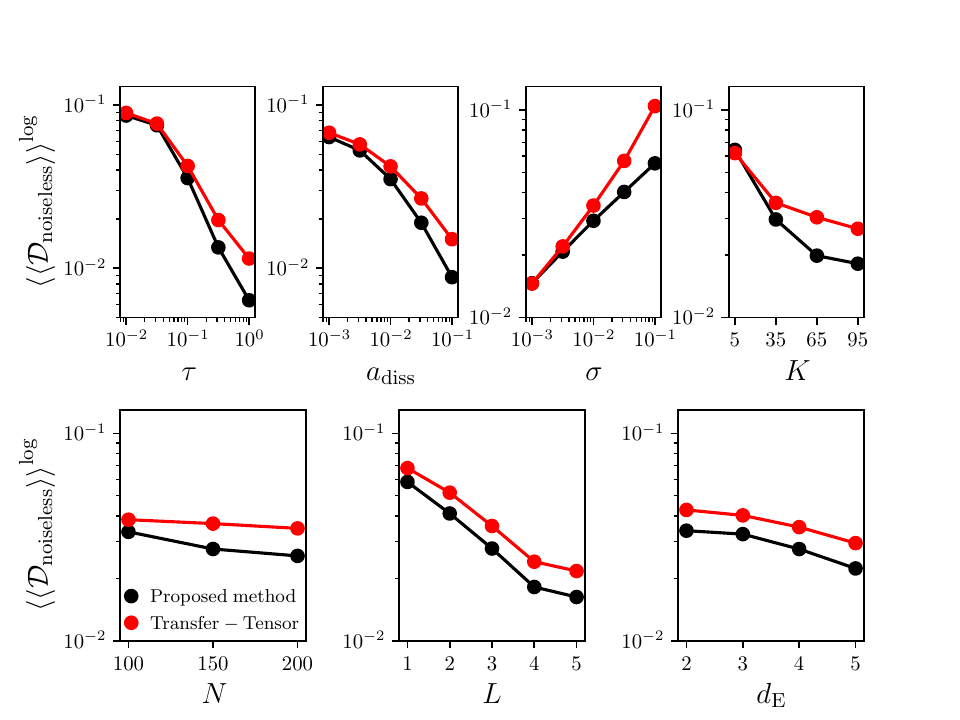}
    \caption{The comparison of the averaged prediction accuracy of the proposed method and the TTM for the model with a predefined effective environment (Eq.~\eqref{eq:model_with_predefined_effective_environment}). The value of ${\cal D}_{\rm noiseless}$ is the distance Eq.~\eqref{eq:dist_between_trajectories} between the prediction built on top of noisy ``prehistory'' of a test trajectory and the noiseless remaining part of a trajectory. The first averaging sign in $\langle\langle {\cal D}_{\rm noiseless}\rangle\rangle^{\rm log}$ means averaging the distance over all trajectories in the test set, the second averaging sign with superscript $\rm log$ means that for each particular plot one performs logarithmic averaging, i.e. $\langle \langle {\cal D}_{\rm noiseless}\rangle \rangle = \exp\left(\langle\log \langle {\cal D}_{\rm noiseless}\rangle \rangle\right)$, over all parameters $\tau, \sigma, K, N, L, d_\rE, a_{\rm diss}, \kappa$ but the parameter that is the $X$-axis.}
    \label{fig:dmd_vs_tt}
\end{figure}

The next indicators of performance are data efficiency and robustness to noise. Our method is equipped with the automatic model selection that selects the simplest possible model among describing observed data. This is achieved by identification of the low-dimensional linear subspace where all the quantum trajectories are placed. The dynamics of the system in this subspace can be described by a model with a fewer number of parameters, which implies less data for learning and better robustness against noise. To demonstrate the data efficiency and robustness to noise of our method, we compare the prediction of a non-Markovian quantum dynamics built using our method with that which is built using the TT method that serves as an example of a ``black-box'' method, and which does not account for the physics under the hood of the observed data. To make our comparison systematic we performed multiple numerical experiments ($300000$ in total) and compared the performance of both methods for each experiment. For the learning of the TTM model from the data, we utilized a ridge regression \cite{bishop2006pattern} to solve the following optimization problem:
\begin{equation}
    \underset{W}{\rm minimize} \ \sum_{i=1}^L\sum_{t=0}^{N-K-1}\left\|\varrho^{(i)}(t+K+1) - W[T_K^{(i)}(t)]\right\|_F^2 + \theta \left\|W\right\|^2_F,
\end{equation}
where $W$ is the kernel of the discrete Nakajima–Zwanzig equation that one reconstructs from the data and $\theta$ is the regularization coefficient taken equal to $\sigma$ in order to mitigate the effect of noise. Each experiment consisted of a data preparation stage, when test and training sets of trajectories are generated within a particular model, a fitting stage, when both the proposed method and the TTM are applied to a training trajectories set in order to build data-driven model of the dynamics, and a test stage, when both data-driven models are used to predict dynamics of trajectories from a test set. After the test stage, prediction errors of both methods are evaluated. As the error measure we used $\left\langle{\cal D}_{\rm noiseless}\right\rangle$ that is distance Eq.~\eqref{eq:dist_between_trajectories} between the prediction built on top of noisy ``prehistory'' of a test trajectory and the noiseless remaining part of a trajectory averaged over all trajectories from a test set. As a model of non-Markovian dynamics we used the model with a predefined effective environment considered in the subsection \ref{seq:Probing non-Markovian quantum dynamics with a predefined effective environment}. Each experiment is defined by a particular tuple of parameters $(\tau, \sigma, K, N, L, d_\rE, a_{\rm diss}, \kappa)$ where $\tau$ is the time step size, $\sigma$ is the amplitude of noise, $K$ is the guess of memory depth, $N$ is the number of discrete time steps in test and training sets, $L$ is the number of trajectories in a training set, $d_\rE$ is the predefined dimension of the effective environment, $a_{\rm diss}$ is the dissipation rate (see Appendix~\ref{appendix:random_linbladians_gen}) and $\kappa$ is the random seed that is used to generate the Lindbladian driving the dynamics of the system and the environment. We performed numerical experiments for all tuples of parameters from the set
\begin{eqnarray}
    &&\left(\tau^{(1)},\dots,\tau^{(5)}\right)\times\left(\sigma^{(1)},\dots, \sigma^{(5)}\right)\times\left(K^{(1)},\dots,K^{(4)}\right)\times\left(N^{(1)},\dots,N^{(3)}\right)\times\left(L^{(1)},\dots,L^{(5)}\right)\times \left(d_\rE^{(1)},\dots, d_\rE^{(4)}\right) \nonumber\\
    &&\times \left(a_{\rm diss}^{(1)},\dots,a_{\rm diss}^{(5)}\right) \times \left(\kappa^{(1)}, \dots, \kappa^{(10)}\right),
\end{eqnarray}
where $\times$ stands for the Cartesian product and superscripts mark a particular value of a parameter used in numerical experiments. In order not to list all values of parameters here we refer a reader to Fig.~\ref{fig:dmd_vs_tt} where $x$-axis of subplots shows the ranges of parameters we used. 

The obtained high-dimensional data needs to be visualized in a concise way. To do that we introduce a logarithmic averaging that we define as follows
\begin{equation}
    \left\langle A \right\rangle^{\log} = \exp\left(\langle\log A\rangle\right),
\end{equation}
that is more preferable than the standard averaging since the value of $\left\langle{\cal D}_{\rm noiseless}\right\rangle$ may vary several orders of magnitude for different parameters. For the sake of demonstration of what is done next, let us fix $\sigma$ as a target parameter. For $\sigma$ we calculated $\left\langle\left\langle{\cal D}_{\rm noiseless}\right\rangle\right\rangle^{\log}_{\tau, K, N, L, d_{\rm E}, a_{\rm diss}, \kappa}$ that is the logarithmic averaged value of $\left\langle{\cal D}_{\rm noiseless}\right\rangle$ over all parameters but $\sigma$. We did the same averaging for all parameters except $\kappa$ and plot the results in Fig.~\ref{fig:dmd_vs_tt}. One can think of this plots as ``projections'' of the obtained high-dimensional data on a particular plane. They represent how the prediction accuracy for both methods depends on a particular parameter. One can note, that the proposed method outperforms the TTM almost always except cases with small noise, small memory depth, small time step size, where both methods perform equally. One can also note, that the proposed method compared to the TTM works especially good in case of high amplitudes of noise, that is especially practical one. These observations approve the proposed method's improved data-efficiency and robustness to noise in comparison with previously propose ``black-box'' methods such as TTM of RNN based method.

To conclude the presented comparison, one can note that the proposed method is a substantial step forward towards interpretable data-driven methods of non-Markovian quantum dynamics analysis. This is the first method allowing automatic extraction of information about the environment from the measured system dynamics and it outperforms previously proposed ``black-box'' methods, such as TTM, in terms of robustness to noise and data efficiency.

\begin{table}[]
\centering
\begin{tabular}{ | p{3cm} | p{3cm}| p{3cm} | p{3cm} | p{3cm} |} 
  \hline
  & Information about the environment that can be reconstructed & Data efficiency and robustness to noise & Mathematical guarantees of convergence to the optimal point & Is able to predict the response of a non-Markovian quantum system on an external perturbation? \\
  \hline
  TTM &  & The most general model requires a large data set  & Converges with mathematical guarantees & No \\
  \hline
  Methods based on learning of a matrix product state representation of a process tensor & Effective dimension \cite{luchnikov2020machine}  & Cross-validation based bond dimension tuning improves data efficiency & No mathematical guarantees of convergence & Yes \\
  \hline
  Methods based on tomography of a process tensor & Eigenfrequencies, effective dimension & Extremely data demanding (requires exponential in the number of discrete time steps amount of data points) & Converges with mathematical guarantees & Yes \\
  \hline
  RNN based method & & "Black box" model requires a large data set & No mathematical guarantees of convergence & No \\
  \hline
  Proposed method & Eigenfrequencies, effective dimension & The automatic model selection improves data efficiency and robustness to noise in contrast with ``black box'' methods& Converges with mathematical guarantees & No \\
  \hline
\end{tabular}
\caption{Comparison of different non-Markovian quantum dynamics identification methods with the method proposed in the present research.}
\label{table:methods_comparison}
\end{table}

\section*{Data availability}
Realization of the method in Python as well as numerical experiments for the model with predefined effective environment are available publicly in https://github.com/LuchnikovI/Automatic-non-markovian-dynamics-learner.

\section{Discussion and outlook}
We developed a data-driven method for non-Markovian quantum dynamics analysis. The input data processed by the method is a set of quantum trajectories, i.e. sequences of density matrices of the system at consecutive time steps. We note that quantum trajectories can be reconstructed with the use of quantum state tomography that is de facto the gold standard for the characterization of quantum information processing devices~\cite{cramer2010efficient}, and is a standard tool for studying the performance of various physical quantum computing platforms including photons~\cite{white1999}, trapped ions~\cite{roos2004}, and superconducting circuits~\cite{berkley2003}. So, the method we developed is well-suited to the study of existing noisy intermediate-scale quantum (NISQ) devices including listed above.
We also note that the developed approach takes into account inevitable discrepancies between true density matrices (which one would obtain with infinite number of perfect measurements) and practically reconstructed ones obtained with finite amount of experimental statistics.
The estimated level of this reconstruction noise (the standard deviation $\sigma$), as well as the guess about the memory depth $K$, are the method's hyperparameters.

The analysis of the non-Markovian dynamics performed within our method consists of two major steps.
At the first stage, the method determines the simplest model describing observed data, in conformity with Occam's razor principle, and outputs the hyperparameter $r$ defining the complexity of the model.
This simplest model selection corresponds to the best machine learning practices and improves both data efficiency and noise robustness of the proposed method \cite{bishop2006pattern}. Moreover, $r$ defines the dimension of the effective environment that is an important characteristic of the genuine environment. 
At the second stage, the method reconstructs the predictive model based on the results at the previous step. 
The obtained model allows one to predict the continuation of the quantum system's trajectory consisting of at least $K$ elements.

Both stages rely on matrix decompositions guaranteeing convergence to the optimal solution while most of previously proposed methods rely on an approximate non-convex optimization \cite{luchnikov2020machine, banchi2018modelling, guo2020tensor, shrapnel2018quantum}. Due to the established relation between a quantum system's trajectories and the joint system and environment dynamics, the reconstructed predictive model is not just a ``black box'', but an interpretable physical model of non-Markovian quantum dynamics.
It allows not only the system's dynamics prediction, but also provides additional information on the underlying open quantum system properties, in particular, the effective dimension of the environment $d^{\rm eff}_\rE$ and eigenfrequencies of the joint system and environment dynamics. 
As an additional valuable output, the proposed method performs quantum trajectories denoising by projecting them onto the $r$-dimensional principle subspace.
We note that the introduced denoising procedure can be used in combination with other approaches for improving accuracy of quantum tomography, including compressive sensing~\cite{gross2010quantum} and employing tensor-product structures~\cite{cramer2010efficient}.

We have validated the performance of our approach for dynamics of two-level system interacting either with a known finite-dimensional effective environment or with an infinite dimensional environment (JC model and spin-boson model).
For the first case, we have justified the ability of the method to recover the correct information on effective environment, and for the latter, we have validated the performance of the method with respect to the models related to actual experiments (see e.g.~\cite{lee2017jc, magazzu2018probing}).
We note that in both cases, the considered values of reconstruction noise $\sigma$, are taken to mimic realistic tomographic experiments.

Our technique is highly relevant for ongoing experiments with NISQ processors, where quantum state tomography of several-qubit systems is an available tool. 
Our method makes it possible to extract relevant information about the environment affecting a quantum processor, which is extremely hard to measure directly, and build its dynamics prediction. 
It potentially allows building new types of data-driven controllers for the next generation of quantum processors taking into account non-Markovian dynamics and using non-Markovianity as a resource.

We also note that the method is directly applicable to the analysis of incomplete data (trajectories of some elements of a density matrix or trajectories of single observables). Indeed, one requires only the linearity of $\phi_K$; it can output either a trajectory of density matrices, a trajectory of certain elements of density matrices, or a trajectory of certain observables. However, in this case, the physical meaning of $d^{\rm eff}_\rE$ is different and requires further analysis. This is in the scope of the next works.

Further research is also required to determine whether our scheme is capable of predicting the response of a non-Markovian system to an external perturbation. Such a prediction capability would open a promising line of research with the development of new data-driven quantum control methods. We also need to study how the proposed approach could be generalized on the case of time dependent interaction between a system and its environment. Specifically, is it possible to capture the dynamics with varying $K$? This generalization would be useful to detect transitions between different dynamics regimes and dynamical phase transitions.

\section*{Acknowledgements}
The development of the data processing scheme, analysis of the spin-boson model, and analysis of the damped Jaynes-Cummings model are supported by the Russian Science Foundation (19-71-10092), by the Leading Research Center on Quantum Computing (Agreement no. 014/20; analysis of non-Markovian processes for NISQ devices), and and by the Priority 2030 program at the National University of Science and Technology ``MISIS” (applications to various quantum models). The analysis of the finite-environment-induced non-Markovian quantum dynamics is supported by the Foundation for the Advancement of Theoretical Physics and Mathematics ``BASIS'' for support under Project No. 19-1-2-66-1.
The authors thank Alexander Ryzhov and Georgiy Semin for fruitful discussions.

\ifx
\section*{Contributions}
I.A.L. came up with the orginal idea of the presented approach.
I.A.L., E.O.K., M.A.G. performed verification of the approach on various physical models of non-Markovian dynamics.
All authors analyzed and interpreted the data.
I.A.L., E.O.K., and A.K.F. wrote the paper with critical input from M.A.G., H.O, and S.N.F.
\fi

\ifx
\section*{Competing interests}
Owing to the employments and consulting activities of I.A.L., E.O.K., and A.K.F., they have financial interests in the commercial applications of quantum computing.
\fi

\appendix
\section{Sufficiency of $K$}
\label{appendix_1}
In order to introduce the criterion of $K$ to be sufficient, i.e. to understand which value of $K$ is enough for the Markovianity of trajectories dynamics, let us properly define the linear map $\phi_K$ that is not discussed in detail in the main text. We define this linear map through its action on an arbitrary joint system and environment state that reads
\begin{equation}
    \phi_K: \varrho_{\rSE} \mapsto \left({\rm Tr}_\rE(\varrho_{\rSE}), {\rm Tr}_\rE(\Phi[\varrho_{\rSE}]), \dots, {\rm Tr}_\rE\left(\Phi^{K-1}[\varrho_{\rSE}]\right)\right).
\end{equation}
The object of our interest is the orthogonal complement of the kernel of $\phi_K$, i.e. the relevant subspace that is introduced in the main text. One can note, that the relevant subspace is essentially an example of a Krylov subspace due to the structure of $\phi_K$. Like for any Krylov subspace its dimension $r(K)$ increases with increasing $K$ until some critical values of $r$ and $K$ are reached after which $r$ saturates and remains constant (see for example \cite{gutknecht2007brief}, Lemma 1). The effect of saturation of $r$ takes place when
\begin{equation}
    \label{eq:orth_kernels_eq}
    {\rm ker}^\perp \left(\phi_{K+1}\right) = {\rm ker}^\perp \left(\phi_{K}\right),
\end{equation}
which means that the relevant subspace does not change with increasing $K$ anymore. More precisely it can be formulated as follows. There exist $K_{\rm c}$ such that $r(K) < r(K+1)$ if $K < K_{\rm c}$ and $r(K) = r(K+1) = r_{\rm c}$ if $K \geq K_{\rm c}$. Let us show that this is indeed true. First of all $r$ is bounded from above, since a linear span of any number of vectors from a finite dimensional linear space cannot be greater that the dimension of this space. Second, $r(K)$ is non-decreasing function of $K$, because by adding new vectors to a linear span one can not discrease the dimension of this linear span. Finally, one needs to prove that once ${\rm ker}^\perp \left(\phi_{K+1}\right) = {\rm ker}^\perp \left(\phi_{K}\right)$, then for any $k\in \mathbb{N}$ the following is true ${\rm ker}^\perp \left(\phi_{K+ k}\right) = {\rm ker}^\perp \left(\phi_{K}\right)$. The equality ${\rm ker}^\perp \left(\phi_{K+1}\right) = {\rm ker}^\perp \left(\phi_{K}\right)$ implies that ${\rm ker}^\perp \left(\phi_{K+1}\Phi\right) = {\rm ker}^\perp \left(\phi_{K}\Phi\right)$ and ${\rm span}\left(\phi_1, {\rm ker}^\perp \left(\phi_{K+1}\Phi\right)\right) = {\rm span}\left(\phi_1, {\rm ker}^\perp \left(\phi_{K}\Phi\right)\right)$. Now note that
\begin{eqnarray}
    &&{\rm span}\left(\phi_1, {\rm ker}^\perp \left(\phi_{K}\Phi\right)\right) = {\rm ker}^\perp(\phi_{K+1}),\nonumber\\
    && {\rm span}\left(\phi_1, {\rm ker}^\perp \left(\phi_{K+1}\Phi\right)\right) = {\rm ker}^\perp(\phi_{K+2}),
\end{eqnarray}
which leads us to the equality ${\rm ker}^\perp \left(\phi_{K+2}\right) = {\rm ker}^\perp \left(\phi_{K+1}\right)$. Applying the same logic again and again we get ${\rm ker}^\perp \left(\phi_{K+3}\right) = {\rm ker}^\perp \left(\phi_{K+2}\right)$, ${\rm ker}^\perp \left(\phi_{K+4}\right) = {\rm ker}^\perp \left(\phi_{K+3}\right)$, etc. This proves the equality ${\rm ker}^\perp \left(\phi_{K+ k}\right) = {\rm ker}^\perp \left(\phi_{K}\right)$. Gathering all together we get the described above behavior, i.e. $r(K)$ strictly increases up to some critical value with increasing $K$ and then saturates and remains constant.

The equality Eq.~\eqref{eq:orth_kernels_eq} is equivalent to the following relation
\begin{equation}
    {\rm ker}^\perp\left(\phi_K\Phi\right)\subseteq {\rm ker}^\perp \left(\phi_K\right).
\end{equation}
This relation implies that there exists such a linear map $M$ that
\begin{equation}
    \phi_K \Phi = M\phi_K,
\end{equation}
which immediately leads to
\begin{equation}
    T_K(t+1) = \phi_K \Phi[\varrho_\rSE(t)] = M\phi_K[\varrho_\rSE(t)] = M[T_K(t)],
\end{equation}
i.e. there exists a Markovian master equation driving the dynamics of trajectories of size $K$ for any initial $\varrho_{\rSE}$. One can note that the inverse statement is also true. This brings us to the criterion of $K$ to be sufficient:
\begin{criterion}
The value of $K$ is sufficient for Markovianity of quantum trajectories iff
\begin{equation}
    {\rm ker}^\perp\left(\phi_K\Phi\right)\subseteq {\rm ker }^\perp \left(\phi_K\right).
\end{equation}
\end{criterion}
Let us derive an important upper bound of the minimal sufficient $K$. Note, that if $K$ is insufficient, then the following holds
\begin{equation}
    r(K+1) \geq r(K) + 1,
\end{equation}
or in other words
\begin{equation}
    r(K) > K.
\end{equation}
Note also that the dimension of the relevant subspace does not exceed $d^2d^2_\rE$. Therefore one has
\begin{equation}
\label{eq:de_r_k_inequality}
    d^2d^2_\rE\geq r(K) > K,
\end{equation}
which states that any insufficient $K$ is less then $d^2d^2_\rE$. This means that the minimal sufficient $K$ is not greater than $d^2d^2_\rE$ and for any finite dimensional environment the minimal sufficient $K$ is also finite. This also means that the minimal sufficient $K$ is less or equal to $r$.

The Eq.~\ref{eq:de_r_k_inequality} implies that in the worst case the minimal sufficient $K$ and $r$ grow exponentially with the number of subsystems of the environment. In the thermodynamic limit the minimal sufficient $K$ and $r$ may even tend to infinity. Our conjecture is that this case corresponds to the dynamical chaos. Indeed, the same unbounded growth of $r$ takes place when one builds time-delay embedding of classical chaotic dynamics \cite{arbabi2017ergodic}. This also is well compatible with understanding of $r$ as the complexity measure. When $r$ and the minimal $K$ tend to infinity this means infinite complexity or chaotic behavior.

\section{GKSL generator of the model with predefined effective environment}\label{appendix:random_linbladians_gen}

In this subsection we describe how we obtain a GKSL generator ${\cal L}:{\cal B}({\cal H} \otimes {\cal H}_\rE) \rightarrow {\cal B}({\cal H} \otimes {\cal H}_\rE)$ driving dynamics of the joint system and effective environment density matrix. The corresponding Lindblad equation reads:
\begin{equation}
    \frac{d\varrho_\rSE}{d t} = {\cal L}(\varrho_\rSE) = -\imath a_{\rm unit}[H, \varrho_\rSE] + a_{\rm diss}\sum_{i,j=1}^{ d^2d_\rE^2-1}\gamma_{ij}\left(F_i \varrho_\rSE F^\dagger_j - \frac{1}{2}\{F^\dagger_j F_i, \varrho_\rSE\}\right).
    \label{eq:random_lindblad_eq}
\end{equation}
Amplitudes $a_{\rm unit}$ and $a_{\rm diss}$ determine contributions of the Hamiltonian and the dissipative parts of the equation into the dynamics. In all experiments we choose $a_{\rm unit}=1$, $a_{\rm diss} = 0.003$, until otherwise stated. The Hamiltonian $H$ is generated randomly as follows:
\begin{equation}
    H = \frac{1}{2}(A + A^\dagger), \ {\rm Re}(A)\sim {\cal N}(0, I), \ {\rm Im}(A)\sim {\cal N}(0, I),
\end{equation}
where ${\cal N}(0, I)$ is the matrix normal distribution with zero mean and the identity covariance matrix.
The positive semi-definite matrix $\gamma$ that characterizes dissipation in the system and environment is also generated randomly. It reads
\begin{equation}
    \gamma = Q \Lambda Q^\dagger,
\end{equation}
where $\Lambda$ is a diagonal matrix with diagonal elements sampled uniformly from the interval $[0, 1]$ and $Q$ is a unitary matrix generated in two steps: 1) one randomly generates a matrix $A$ such that ${\rm Re}(A)\sim {\cal N}(0, I)$ and ${\rm Im}(A) \sim {\cal N}(0, I)$, 2) one performs QR factorization of $A$ and takes unitary factor as $Q$. The set of real, traceless matrices $\{F_1,F_2,\dots,F_{d^2d_\rE^2-1}\}$ forms an orthonormal basis, i.e. ${\rm Tr}(F_i^\dagger F_j) = \delta_{ij}$, where $\delta_{ij}$ stands for the Kronecker's symbol.

\section{NumPy notations for tensors slicing}
\label{appendix:slicing}
\begin{figure}[h]
    \centering
    \vskip -4mm
    \includegraphics[scale=0.8]{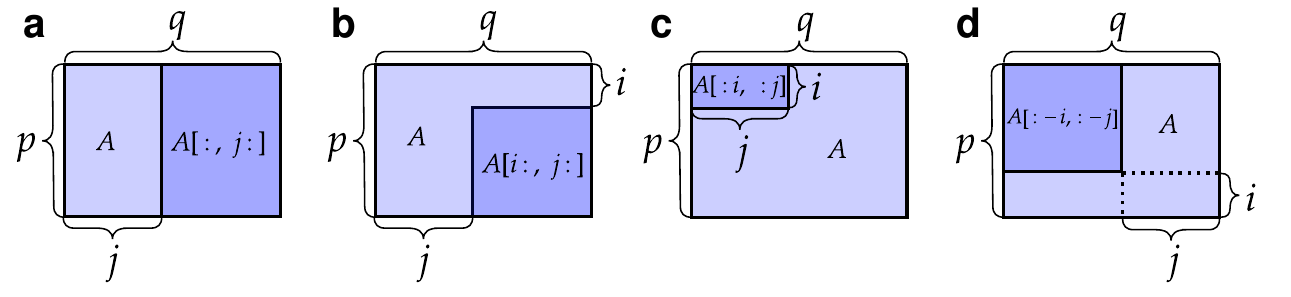}
    \caption{Illustration of NumPy notations for matrices slicing. {\bf a} First $j$ columns are removed, {\bf b} first $j$ columns and first $i$ rows are removed, {\bf c} only first $j$ columns and $i$ rows are kept, {\bf d} last $j$ columns and last $i$ rows are removed.}
    \label{fig:slicing}
\end{figure}
In order to represent submatrices of a matrix we use notations that are standard in many programming packages for numerical computation such as NumPy \cite{harris2020array}. Let us consider a rectangular matrix $A$ of the size $p\times q$, that can be of complex, real, or any other type. If we want to consider a truncated version of $A$ for which we remove first $j$ columns, we write $A[:,j:]$. If we additionally want to remove first $i$ rows, we write $A[i:,j:]$. If we want to keep only first $j$ columns and first $i$ rows, we write $A[:i,:j]$. It is also possible to count columns and rows starting from the other end using negative integers as indices. For example, if we want to remove last $j$ columns and last $i$ rows, we write $A[:-i,:-j]$. All the examples are also illustrated in Fig.~\ref{fig:slicing}.

\bibliography{bibliography.bib}

\end{document}